\documentclass[10pt]{article}

\usepackage[square,numbers]{natbib}
\usepackage{graphicx}
\usepackage[T1]{fontenc}
\usepackage[utf8]{inputenc}
\usepackage{authblk}
\usepackage{fancyvrb} 
\usepackage{color} 
\usepackage{setspace} 
\usepackage{fullpage}
\usepackage{listings}
\usepackage{caption}
\usepackage{subcaption}
\usepackage[]{algorithm2e}
\usepackage{multirow}
\usepackage{bm}
\usepackage{amsfonts}
\usepackage{mathtools}
\usepackage{float}
\usepackage{verbatim}
\usepackage{float}

\title{Prepaid parameter estimation without likelihoods}

\author[1,*,+]{Merijn Mestdagh}
\author[1,+]{Stijn Verdonck}
\author[1]{Kristof Meers}
\author[1]{Tim Loossens}
\author[1]{Francis Tuerlinckx}

\affil[1]{KU Leuven -- University of Leuven, Belgium}
\affil[*]{merijn.mestdagh@kuleuven.be}
\affil[+]{These authors contributed equally to this work.}

\begin{document}

\flushbottom
\maketitle

\thispagestyle{empty}

\begin{abstract}
In various fields, statistical models of interest are analytically intractable. As a result, statistical inference is greatly hampered by computational constraints. However, given a model, different users with different data are likely to perform similar computations. Computations done by one user are potentially useful for other users with different data sets. We propose a pooling of resources across researchers to capitalize on this. More specifically, we preemptively chart out the entire space of possible model outcomes in a prepaid database. Using advanced interpolation techniques, any individual estimation problem can now be solved on the spot. The prepaid method can easily accommodate different priors as well as constraints on the parameters. We created prepaid databases for three challenging models and demonstrate how they can be distributed through an online parameter estimation service. Our method outperforms state-of-the-art estimation techniques in both speed (with a 23,000 to 100,000-fold speed up) and accuracy, and is able to handle previously quasi inestimable models.
\end{abstract}

\section{Author Summary}
Interesting nonlinear models are often analytically intractable. As a result, statistical inference has to rely on massive, time-intensive, simulations. The main idea of our method is to avoid the redundancy of similar computations that typically occur when different researchers independently fit the same model to their particular dataset. Instead, we propose to pool computational resources across the researchers interested in any given model. The prepaid method starts with an extensive simulation of datasets across the parameter space. The simulated data are compressed into summary statistics, and the relation to the parameters is learned using machine learning techniques. This results in a parameter estimation machine that produces accurate estimates very quickly (a 23,000 to 100,000-fold speed up compared to traditional methods).

\section{Introduction}

Models without an analytical likelihood are increasingly used in various disciplines, such as genetics \cite{beaumont_approximate_2002}, ecology \cite{wood_statistical_2010,fasiolo_comparison_2016}, economics \cite{mcfadden_method_1989,fermanian_nonparametric_2004} and neuroscience \cite{turner_bayesian_2016}. For models without an analytical or easily computable likelihood, parameter estimation is a major challenge for which a variety of solutions have been proposed \cite{wood_statistical_2010,beaumont_approximate_2002,heard_agent-based_2015}. All these methods have in common that they rely on extensive Monte Carlo simulations and that their convergence can be painstakingly slow. As a result, the current methods can be very time consuming.

To date, the practice is to analyze each data set separately. However, considering all the calculations that have ever been performed during parameter estimation of a particular type of model, for each different data set, one cannot help but notice an incredible waste of resources. Indeed, simulations performed while estimating one data set may also be relevant for the estimation of another. Currently, each researcher estimating the same model with different data will start from scratch, and not benefit from all the possibly relevant calculations that have already been performed in earlier analyses by other researchers, in other locations, on different hardware, and for other data sets, but concerning the same model.

Therefore, our proposal is to combine resources and create a giant online database of parameter values coupled to simulated data. Consequently, (cloud based) interpolation techniques and global optimization methods can be used on the previously created (hence, prepaid) database for accurate and fast parameter estimation on any device. Statistical analyses that currently take up hours to days of computation time on dedicated hardware are now available to everyone within seconds.

In Figure~\ref{fig:method} we present a graphical illustration of the prepaid parameter estimation method. First (panel A), for a representative number of parameter vectors $\boldsymbol{\theta}$, large data sets are simulated, compressed into summary statistics (i.e., $\boldsymbol{s}^{\mathrm{sim}}$) and saved --- creating the prepaid grid. This prepaid grid is computed beforehand and the results are stored at a central location. Second (panel B1), the observed (data) summary statistics ($\boldsymbol{s}^{\mathrm{obs}}$) are compared to the simulated (data) summary statistics  (i.e., $\boldsymbol{s}^{\mathrm{sim}}$) using an appropriate objective loss function $d\left(\boldsymbol{s}^{\mathrm{sim}},\boldsymbol{s}^{\mathrm{obs}}\right)$ and a number of nearest neighbor simulated summary statistics are selected. The loss function is related to the loss function used in the generalized method of moments \cite{hall_generalized_2005} and method of simulated moments \cite{gourieroux_simulation-based_1996}.

Third (panel B2), interpolation methods are used to find the relation $\boldsymbol{s}=f(\boldsymbol{\theta})$ between the parameter values and the summary statistics for the selected points of the previous step \cite{gutmann_bayesian_2016, mestdagh_fingerprint_2015}. In this paper, we use tuned least squares support vector machines, LS-SVM \cite{suykens_least_2002}. Finally (panel B3), the objective loss function $d\left(\boldsymbol{s}^{\mathrm{pred}},\boldsymbol{s}^{\mathrm{obs}}\right)$, now using predicted summary statistics $\boldsymbol{s}^{\mathrm{pred}}$, is minimized as a function of the unknown parameter values using an optimizer.

Three important aspects of the prepaid method deserve special mention. First, the parameter space is required to be bounded. If this is unnatural for a given parametrization, then the parameters have to be appropriately transformed to a bounded space. Second, we typically start from a uniform distribution of parameter vectors in the final parameter space. This choice reflects on the uniformity of the grid's resolution, but has no further implications provided the grid is sufficiently dense. Bayesian priors can be implemented without recreating the prepaid grid, since the prior can be taken into account in the loss function. Third, often a user is not interested in a single instance of a model, but rather has data from several experimental conditions that share some common parameters but assume other ones to be different. Also in these cases the prepaid grid does not need to be recreated, as the parameter constraints can be included through priors with tuning parameters (i.e., penalties).

The performance of the prepaid method can be studied theoretically in simple situations (see Methods~\ref{subs:toy}). In what follows, the prepaid method will be applied to three more complicated, realistic scenarios.

\newcommand{\widthpercentagepicture}{0.7} %percentage of textwidth
\newcommand{\heightdivweight}{1.7798} %ratio height/width of figure
\newsavebox{\mybox} 
\savebox{\mybox}{\includegraphics[width=\widthpercentagepicture\textwidth]{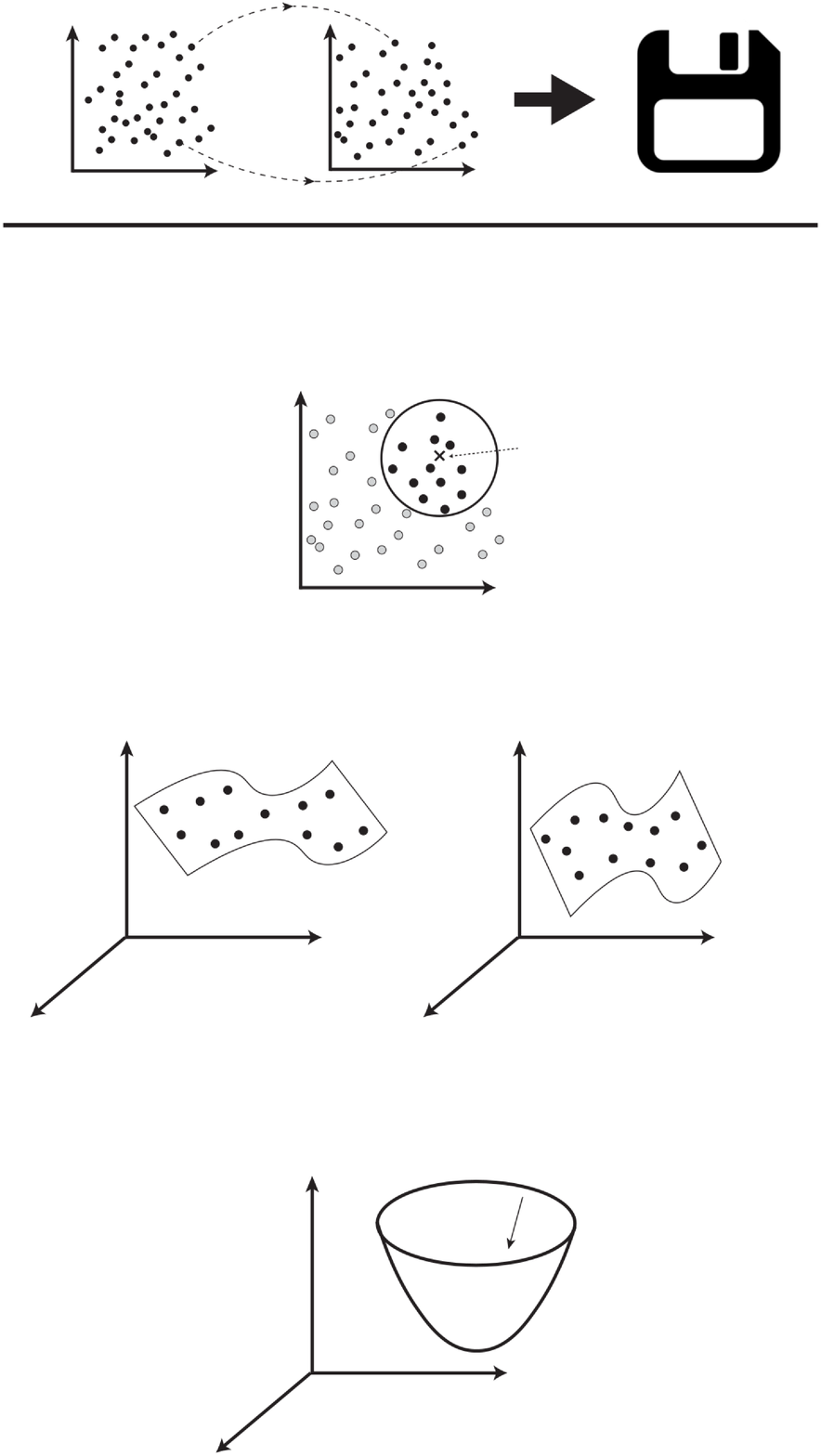}} 
\setlength{\unitlength}{\widthpercentagepicture\textwidth}

\begin{figure}[H]
\centering
%\begin{picture}(1,\heightdivweight)
\begin{picture}(1,2)
\put(0,0){\usebox{\mybox}}
%TITLES
\put(0,1.83){\textbf{A) Prepaid}}
\put(0.06,1.77){Simulate and Save}
\put(0,1.37){\textbf{B) Estimation} }
\put(0.06,1.31){1) Nearest Neighbors}
\put(0.06,0.93){2) Local Learning}
\put(0.06,0.43){3) Method of Predicted Moments}
% simulate and save
\put(0.08,1.64){$\theta_2$}
\put(0.335,1.64){$s_2^{\mathrm{sim}}$}
\put(0.26,1.463){$\theta_1$}
\put(0.53,1.463){$s_1^{\mathrm{sim}}$}

% nearest neighbors
\put(0.322,1.219){$s_2$}
\put(0.576,0.974){$s_1$}
\put(0.645,1.17){$\boldsymbol{s^{\mathrm{obs}}}=\left\{s^{\mathrm{obs}}_1,s^{\mathrm{obs}}_2\right\}$}
\put(0.545,1.243){$d\left(\boldsymbol{s^{\mathrm{sim}}},\boldsymbol{s^{\mathrm{obs}}}\right) \leq \mbox{constant}$}

% local leraning
\put(0.14,0.84){$s_1$}
\put(0.38,0.582){$\theta_2$}
\put(0.03,0.548){$\theta_1$}
\put(0.24,0.85){$s_2=f_1\left(\theta_1,\theta_2\right)$}

\put(0.582,0.84){$s_2$}
\put(0.862,0.582){$\theta_2$}
\put(0.492,0.548){$\theta_1$}
\put(0.72,0.85){$s_2=f_2\left(\theta_1,\theta_2\right)$}

% method of predicted moments
\put(0.312,0.37){$d\left(\boldsymbol{s}^{\mathrm{pred}},\boldsymbol{s}^{\mathrm{obs}} \right)= d\left(\left\{f_1\left(\theta_1,\theta_2\right),f_2\left(\theta_1,\theta_2\right)\right\},\boldsymbol{s^{\mathrm{obs}}}\right)$}
\put(0.592,0.075){$\theta_2$}
\put(0.232,0.033){$\theta_1$}

\end{picture}
\caption{Graphical illustration of the prepaid parameter estimation method.}
\label{fig:method}
\end{figure}

\section{Results}

\subsubsection*{Example 1: The Ricker model}
In a first example, we apply our prepaid method to the Ricker model \cite{turchin_complex_2003,wood_statistical_2010} which describes the dynamics of the number of individuals $y_t$ in a species over time (with $t=1$ to $T_{\mathrm{obs}}$):
\begin{equation} \label{eqRicker}
\begin{split}
 y_{t}& \sim \text{Pois}(\phi N_t) \\
N_{t+1}&=rN_{t}\operatorname{e}^{-N_{t}+e_{t}}
\end{split}
\end{equation}
where $e_t \sim \mathcal{N}\left(0,\sigma^2\right)$. The variables $N_t$ (i.e., the expected number of individuals at time $t$) and $e_t$ are hidden states. Given an observed time series $\{y_t\}_{t=1}^{T_{\mathrm{obs}}}$, we want to estimate the parameters $\boldsymbol{\theta}=\left\{r,\sigma,\phi\right\}$, where $r$ is the growth rate, $\sigma$ the process noise and $\phi$ a scaling parameter. The Ricker model can demonstrate near-chaotic or chaotic behavior and no explicit likelihood formula is available. 

Wood \cite{wood_statistical_2010} used the synthetic likelihood to estimate the model's parameters. In the original synthetic likelihood approach (denoted as $\text{SL}^\text{Orig}$), the assumed multivariate normal distribution of the summary statistics is used to create a synthetic likelihood. The mean and covariance matrix of this normal distribution are functions of the unknown parameters and are calculated using a large number of model simulations. The synthetic likelihood is proportional to the posterior distribution from which is sampled using MCMC and a posterior mean is computed.

Wood's synthetic likelihood $\text{SL}^\text{Orig}$ approach is compared to the prepaid method, where we create a prepaid grid of the mean and the covariance matrix of a similar set of summary statistics. Prepaid estimation comes in multiple variants, depending on the use of an interpolation method. The first, which uses only the prepaid grid points and chooses the nearest neighbor (maximum synthetic likelihood) as final estimate, will be called $\text{SL}^{\text{Grid}}_{\text{ML}}$. The second, $\text{SL}^{\text{SVM}}_{\text{ML}}$, uses LS-SVM to interpolate between the parameters in the prepaid grid to increase accuracy. The differential evolution algorithm (a global optimizer; \cite{storn_differential_1997}) is used to maximize this interpolated synthetic (log)likelihood. Additional details on the implementation of the synthetic likelihood can also be found in Methods~\ref{subs:ricker}.

\begin{figure}[H]
  \centering
    \includegraphics[width=0.5\textwidth]{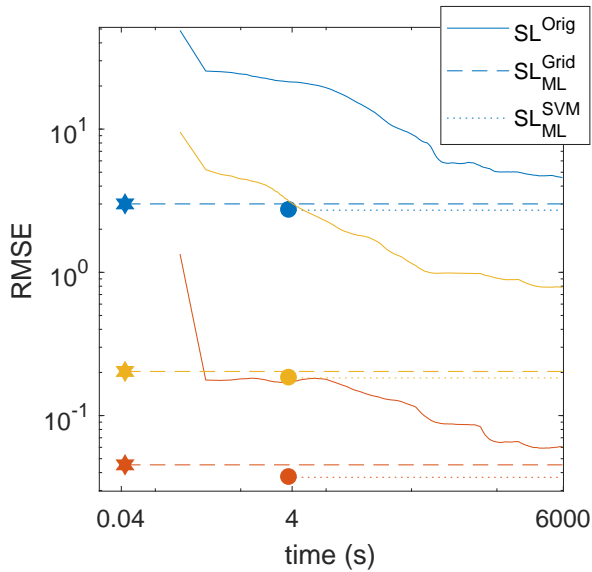}
     \caption{The RMSE versus the time needed for the estimation of the three parameters of the Ricker model (see Equation~\ref{eqRicker}). The RMSE and time are based on 100 test data sets with $T_{\mathrm{obs}}=1000$. The three colors represent the three parameters (blue for $r$, red for $\sigma$ and yellow for $\phi$). Solid lines represent the $\text{SL}^\text{Orig}$ approach, dashed lines the $\text{SL}^{\text{Grid}}_{\text{ML}}$ approach (using only nearest neighbors), and dotted lines the $\text{SL}^{\text{SVM}}_{\text{ML}}$ approach (using interpolation). The stars and the dots represent the time needed for the $\text{SL}^{\text{Grid}}_{\text{ML}}$ and the $\text{SL}^{\text{SVM}}_{\text{ML}}$ estimation, respectively. The estimates for $\text{SL}^\text{Orig}$ are posterior means, based on the second half of the finished MCMC iterations.}
     \label{rickert}
\end{figure}

Figure~\ref{rickert} shows both the accuracy of parameter recovery (as measured with the RMSE) and computation time for the three methods under comparison: (1) $\text{SL}^\text{Orig}$ as in \cite{wood_statistical_2010}, the prepaid method (2) with ($\text{SL}^{\text{SVM}}_{\text{ML}}$), and (3) without ($\text{SL}^{\text{Grid}}_{\text{ML}}$) interpolation. As can be seen in Figure~\ref{rickert}, the prepaid estimation techniques lead to better results than the synthetic likelihood for $T_{\mathrm{obs}}=1,000$, both in accuracy and speed. The $\text{SL}^\text{Orig}$ method leads to some clear outliers (see Methods~\ref{subs:ricker} ) which testifies to possible convergence problems (probably due to local minima). The prepaid method suffers much less from this problem. Most striking is the speed up of the prepaid method: The  $\text{SL}^{\text{Grid}}_{\text{ML}}$ version of the prepaid estimation is finished before a single iteration of the 30,000 iterations in the synthetic likelihood method has been completed --- 100,000 times faster. In addition, it is demonstrated that the coverages of the prepaid method confidence intervals are very close or exactly equal to the nominal value (we look at 95\% bootstrap-based confidence intervals). SVM interpolation is mainly helpful for large $T_{\text{obs}}$, where one expects a higher accuracy of the estimates and the grid is too coarse. The analyses with large $T_{\text{obs}}$ could only be completed in a reasonable time using the prepaid method (See Methods~\ref{subs:ricker} for more detailed information).

In the application above, the tacitly assumed prior on the parameter space is uniform. In addition, there is only one data set for which a single triplet of parameters $(r,\sigma,\phi)$ needs to be estimated. In Methods~\ref{subs:ricker}, we show how both limitations can be relaxed. First, it is explained how different priors for the Ricker model can be implemented. Second, it is discussed what can be done if there are two data sets (i.e., conditions) for which it holds that $r_1=r_2$ and $\sigma_1=\sigma_2$ but $\phi_1$ and $\phi_2$ are not related.

Finally, we also tested our estimation process on the population dynamics of the Chilo partellus, extracted from Figure 1 in Taneja and Leuschner \cite{taneja_methods_1985, yonow_potential_2017}. Here we found that $r=1.10$ (95\% confidence interval 1.06-- 1.34), $\sigma=0.43$ (95\% confidence interval 0.30 -- 0.54) and $\phi=140.60$ (95\% confidence interval = 43.94 -- 208.19). We found similar results  using the synthetic likelihood method (see Methods ~\ref{subs:ricker}), but our estimation was 4000 times faster.

%The  $\text{SL}^{\text{Grid}}_{\text{ML}}$ version of the prepaid estimation is finished %before a single iteration of the 30,000 iterations in the synthetic likelihood method has %been completed --- 100,000 times faster. 
%FT: bootstrap-based added.

\subsubsection*{Example 2: A stochastic model of community dynamics}

A second example we use to illustrate the prepaid inference method is a trait model of community dynamics \cite{jabot_stochastic_2010} used to model the dispersion of species. For this model (see also Methods section), there are four parameters to be estimated: $I$, $A$, $h$, and $\sigma$. As with the first application, there is no analytical expression for the likelihood \cite{jabot_stochastic_2010}. 

As an established benchmark procedure for this trait model, we apply the widely used Approximate Bayesian Computation (ABC) method \cite{csillery_approximate_2010,voight_bayesian_2012,siepel_bayesian_2011,beaumont_approximate_2010} as implemented in the Easy ABC package and denoted here as $\text{ABC}^{\text{Orig}}_{\text{PM}}$ (PM stands for posterior means, which will be used as point estimates) \cite{jabot_easyabc:_2015}. As priors, we use uniform distributions on bounded intervals for $\log(I)$, $\log(A)$, $h$ and $\log(\sigma)$ (see Methods~\ref{subs:comdyn} for the exact specifications), but this can be easily changed as explained for the first example.

To allow for a direct comparison with the ABC method, and to illustrate the versatility of the prepaid method, we have also implemented three Bayesian versions of the prepaid method. The first, $\text{SL}^{\text{Grid}}_{\text{PM}}$, creates a posterior proportional to the prepaid synthetic likelihood. The second, $\text{ABC}^{\text{Grid}}_{\text{PM}}$, saves not only, the mean and covariance matrix of the summary statistics for every parameter in the prepaid grid, but also a large set of uncompressed summary statistics. Using these statistics we are able to approximate an ABC approach. The third, $\text{ABC}^{\text{SVM}}_{\text{PM}}$, again interpolates between the grid points to achieve a higher accuracy. 
%FT WAAROM TABLE4 UIT SUP MATS HIER NIET LATEN ZIEN? EVT MET KOLOM TOEGEVOEGD OVER SPEED

\begin{table}[H]
\caption{The RMSE of the estimates of the test set of the trait model. $T_{\mbox{obs}}$ refers to the number of observations (i.e., vector with species frequencies) and $\Omega$ is the number of prepaid points. }
\begin{center}
\begin{tabular}{c l l| r r r r}
 $T_{\mathrm{obs}}$ & version & $\Omega$ & $\log (I)$ & $\log(A) $ & $h$  & $\log(\sigma)$\\ \hline
  1 & $\text{ABC}^{\text{Orig}}$ & / & 0.17 & 0.67 & 7.45 & 0.74  \\
  1 & $\text{SL}^{\text{Grid}}_{\text{PM}}$ & 100000   & 0.17 & 0.66 & 7.49 & 0.7\\
  1 & $\text{ABC}^{\text{Grid}}_{\text{PM}}$ & 100000   &  0.16 & 0.63 & 7.9 & 0.7 \\
  1 & $\text{ABC}^{\text{Grid}}_{\text{PM}}$ & 500000   &  0.16 & 0.62 & 8.17 & 0.7 \\ \hline
1000 & $\text{ABC}^{\text{Grid}}_{\text{PM}}$ & 100000  & 0.07 & 0.35 & 6.41 & 0.61 \\
1000 & $\text{ABC}^{\text{Grid}}_{\text{PM}}$ & 500000   & 0.05 & 0.27 & 4.83 & 0.48  \\
1000 & $\text{ABC}^{\text{SVM}}_{\text{PM}}$ & 100000   & 0.03 & 0.23 & 5.24 & 0.42 \\
1000 & $\text{ABC}^{\text{SVM}}_{\text{PM}}$ & 500000  & 0.03 & 0.21 & 4.39 & 0.4 \\
\end{tabular}
\end{center}
\label{tabap2rmse}
\end{table}

All methods result in accuracies of the same order of magnitude as can be seen in Table~\ref{tabap2rmse}.  The main difference is again the speed of the methods: $\text{ABC}^{\text{Grid}}_{\text{PM}}$ is about 23,000 times faster than traditional ABC. For small sample sizes, all ABC based methods achieve good coverage. However, for large sample sizes, $\text{ABC}^{\text{Orig}}_{\text{PM}}$ cannot be used anymore (because of the unduly long computation time). For the prepaid versions, it is necessary to use SVM interpolation between the grid points to get accurate results. 

\subsubsection*{Example 3: The Leaky Competing Accumulator for choice response times}

In a third example, we apply our method to stochastic accumulation models for elementary decision making. In this paradigm, a person has to choose, as quickly and accurately as possible, the correct response given a stimulus (e.g., is a collection of points moving to the left or to the right). Task difficulty is manipulated by applying different levels of stimulus ambiguity.

A popular neurally inspired model of decision making is the Leaky Competing Accumulator (LCA\cite{usher_time_2001}). For two response options, two noisy evidence accumulators (stochastic differential equations, see Methods section) race each other until one of them reaches the required amount of evidence for the corresponding option to be chosen. The time that is required to reach that option's threshold is interpreted as the associated choice response time. For different levels of stimulus difficulty, the model produces different levels of accuracy and choice response time distributions. The evidence accumulation process leading up to these choices and response times is assumed to be indicative of the activation levels of neural populations involved in the decision making.

As in the first two examples, there is no analytical likelihood available that can be used to estimate the parameters of the LCA. Moreover, the LCA is an extremely difficult model to estimate. To the best of our knowledge, only \cite{miletic_parameter_2017} systematically investigated the recovery of the LCA parameters, but for a slightly different model (with three choice options) and with a method that is impractically slow for very large sample sizes, making it difficult to show near-asymptotic recovery properties with.

\begin{figure}[H]
  \centering
    \includegraphics[width=0.5\textwidth]{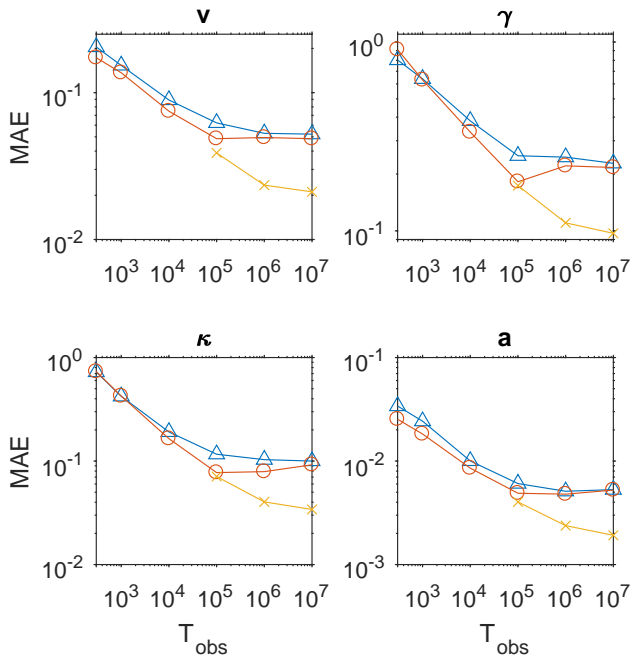}
     \caption{The mean absolute error of the estimates of four central parameters of the LCA (common input $v$, leakage $\gamma$, mutual inhibition $\kappa$, evidence threshold $a$) as a function of sample size (abscissa) and for three different methods: (1) choosing the nearest neighbor grid point in the space of summary statistics ($\text{CHISQ}^{\text{Grid}}_{\text{NN}}$, triangles); (2) using the average of a set of nearest neighbor grid points based on bootstrap samples ($\text{CHISQ}^{\text{Grid}}_{\text{BS}}$, open circles) and (3) using SVM interpolation between the 100 nearest neighbors ($\text{CHISQ}^{\text{SVM}}_{\text{BS}}$, crosses).}
     \label{mae}
\end{figure}

For an experiment with four stimulus difficulty levels, the LCA model has nine parameters. However, after a reparametrization of the model (but without a reduction in complexity), it is possible to reduce the prepaid space to four dimensions (see Methods~\ref{subs:lca}) and conditionally estimate the remaining subset of the parameters with a less computationally intensive method. Three variants of the prepaid method have been implemented: taking the nearest neighboring parameter set (based on a symmetrized $\chi^2$ distance between distributions) on the prepaid grid ($\text{CHISQ}^{\text{Grid}}_{\text{NN}}$), averaging over the grids nearest neighboring parameter sets of 100 non-parametric bootstrap samples ($\text{CHISQ}^{\text{Grid}}_{\text{BS}}$), using SVM interpolation for every bootstrap estimate ($\text{CHISQ}^{\text{SVM}}_{\text{BS}}$). A nearest neighbor or bootstrap averaged estimate completes in about a second on a Dell Precision T3600 (4 cores at 3.60GHz), an SVM interpolated estimate requires a couple of minutes extra.

\begin{figure}[H]
  \centering
    \includegraphics[width=0.5\textwidth]{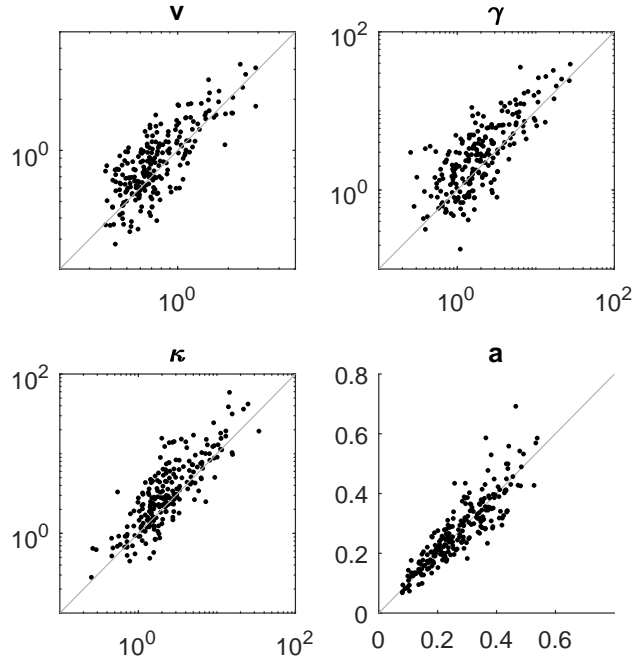}
     \caption{Parameter recovery for the LCA model with 1200 observations ($300$ in each of the four difficulty conditions); the true value on the abscissa and estimated value on the ordinate. The same parameters as in Figure~\ref{mae} are shown. The method used to produce these estimates is the averaged bootstrap approach ($\text{CHISQ}^{\text{Grid}}_{\text{BS}}$, see Methods~\ref{subs:lca} for details).}
     \label{N300}
\end{figure}

Figure~\ref{mae} displays the mean absolute error (MAE) of the estimates for four of the nine parameters as a function of sample size, separately for three estimation methods. The results for the other parameters are similar and can be consulted in the Methods section. It can be seen that with increasing sample size, MAE decreases. The SVM method pays off especially for larger samples. Figure~\ref{N300} shows detailed recovery scatter plots for a subset of the parameters for 1,200 observed trials, which is the typical size of decision experiments. To get better recovery, larger sample sizes have to be considered (see Methods section). In general, recovery is much better than what has been reported in \cite{miletic_parameter_2017}. The coverage of the method, based on non-parametric bootstrapping, is satisfactory for all sample sizes, provided SVM interpolated estimates are used for $T_{obs}>100000$. In addition, we do not find evidence for a fundamental identification issue with the two option LCA, as has been stated in \cite{miletic_parameter_2017}.

\section{Discussion}
In three examples, we have demonstrated the efficacy and versatility of the prepaid method. The  prepaid method is at least as accurate as current methods, but many times faster (23,000 to 100,000-fold speed up). Besides the improvements at the level of speed and accuracy, the prepaid method has a number of other distinct advantages. First, the prepaid method can be used for a very large number of observations, contrary to the synthetic likelihood or ABC methods. The use of very large simulated data sets allows investigation of large-sample properties of the estimator, which is a problem for the synthetic likelihood and ABC. Second, because of the enormous speed improvement and having data sets available across the whole parameter space, the prepaid method allows for fast yet extensive testing of recovery of simulated data across this space --- the recovery of every single parameter set can be evaluated. Such a practice leads to detailed internal quality control of the used estimation algorithm. 

Although the idea behind the prepaid method is fairly simple, we want to anticipate a few misconceptions that might arise. First, as has been demonstrated in the context of the Ricker model, the prepaid method can easily deal with different priors and with equality constraints on parameters, without the need to recreate the underlying prepaid grid. Second, the observed data based on which the model parameters have to be estimated can be of any size, again without the need to recreate the prepaid grid for each and every sample size.

Ideally, the prepaid databases and the corresponding estimation algorithms will be constructed and made available by a team of experts for the model at hand. Subsequently, a cloud based service can then be set up to offer high quality model estimations to a broad public of researchers. As an example, we created such a service for the Ricker model in Equation~\ref{eqRicker}: www.prepaidestimation.org, where we allow the user to estimate the parameters of the Ricker model for personal data as well as 4 example data sets including one real life data set \cite{taneja_methods_1985, yonow_potential_2017}. By using such a cloud based service, researchers that need their data analyzed with computationally challenging models, can avoid many of the pitfalls they would otherwise encounter venturing out on their own. This practice will also lead to increased reproducibility of computational results.  

A first possible objection against the prepaid method is the considerable initial simulation cost (for the examples discussed, prepaid simulations took up to a couple of days on a 20-core processor). However, this overhead cost will dissipate entirely as increasingly more estimates are sourced from the same prepaid database. Moreover, the initial prepaid cost can be easily distributed across multiple interested parties. Further, because the database can be used for internal quality control, additional simulation studies investigating the recovery of parameters are made redundant. 

A second possible objection is that the prepaid grid, unsurprisingly, does not escape the curse of dimensionality: The grid size grows exponentially with the number of parameters. The prepaid method is most effective for highly nonlinear models with substantively meaningful parameters, as they appear in various computational modeling fields. Thus, the number of parameters cannot be very large. However, this limitation can be alleviated in a number of ways. First, the use of interpolation techniques allows for a substantial reduction of the number of prepaid points (by a factor of five for the same accuracy in the trait model example; see Methods section). Second, as is shown in the LCA example, it is possible to only partially apply the prepaid method, estimating with normal techniques, the less challenging parameters conditionally on a prepaid grid of the more intricately connected ones. Third, as shown by tackling three challenging examples, current storage and throughput possibilities can accommodate realistically sized prepaid databases.

It is our strong belief that this method will massively democratize the use of many computationally expensive models, which are now reserved for people with access to specific high-end hardware (e.g., GPUs, HPC). Apart from such democratization, this approach could significantly impact the current work flow of scientific modeling, in which every part of the estimation is carried out locally by an individual researcher. 

% \begin{enumerate}

% \item identification? : how do we know that our summary statistics identify the parameters
% \item related issue is choice of moments: there is a general theory on indirect inference (e.g., using the parameters of an auxiliary model with easy to estimate parameters as moments)
% \end{enumerate}

\section{Methods}

\subsection{A toy example: Estimating the mean of a normal} \label{subs:toy}
For a very simple setting, we want to study the performance of the prepaid methods analytically.

Assume $y_i \sim N(\mu,s^2)$ ($i=1,\dots,T_{\mathrm{obs}}$) with the mean $\mu$ unknown (and to be estimated and the standard deviation $s$ known (so number of parameters $K=1$). The observed mean is denoted as $\bar{y}$. We will explore two situations. In the first situation, $\bar{y}$ will be our summary statistic $s^{\mathrm{obs}}$ (hence number of summary statistics $R=1$) to estimate $\mu$ ($\bar{y}$ is also a sufficient statistic for $\mu$). In the second situation, we will study what happens if  $s^{\mathrm{obs}}= \bar{y}^2$ is chosen to be the summary statistic.

\paragraph{Situation 1: $s^{\mathrm{obs}}= \bar{y}$} 
As a prepaid grid, we take $N_r$ evenly spaced $\mu$-values with spacing or gap size $\Delta=\mu_{j+1}-\mu_j$. For each value $\mu_j$, $T_{\mathrm{sim}}$ values of $y$ are simulated and the sample average is computed (i.e., $\bar{y}^{\mathrm{sim}}_j$). Typically, $T_{\mathrm{sim}}=1000$ or larger. Hence, every value of $\mu_j$ is paired with a particular $\bar{y}^{\mathrm{sim}}_j$: $(\mu_j,\bar{y}^{\mathrm{sim}}_j)$.

Given an observed $\bar{y}$, the $N$ nearest neighbors of simulated statistics $\bar{y}^{\mathrm{sim}}_j$ are selected: $(\mu_{(1)},\bar{y}^{\mathrm{sim}}_{(1)})$, $(\mu_{(2)},\bar{y}^{\mathrm{sim}}_{(2)})$, $\dots$, $(\mu_{(N)},\bar{y}^{\mathrm{sim}}_{(N)})$, such that $|\bar{y}^{\mathrm{sim}}_{(1)}-\bar{y}| \leq |\bar{y}^{\mathrm{sim}}_{(2)}-\bar{y}| \leq \dots \leq |\bar{y}^{\mathrm{sim}}_{(N)}-\bar{y}|$. Typically, $N=100$.

Because of the linearity of the problem, we can safely assume that if $T_{\mathrm{sim}}$ is large enough, the $N$ selected $\mu$ values are all consecutive or nearly consecutive (because of noise in the prepaid simulation of $\bar{y}^{\mathrm{sim}}$, it can happen that the $N$ selected $\mu$ values are not consecutive). We denote the average of these $N$ $\mu$-values as $M_{\mu}$. If all values are exactly consecutive, $M_{\mu}$ can be expressed as

\begin{align*}
M_{\mu} &= \frac{1}{N} \, \sum_{j=1}^{N} \, \mu_{(j)} \\
&= \frac{1}{N} \, \sum_{j=0}^{N-1} \, \left( \mu_{(1)} + j \, \Delta \right) \\
&= \mu_{(1)} + \frac{\Delta}{N} \, \sum_{j=1}^{N-1} \, j \\
&= \mu_{(1)} + \frac{\Delta \left(N-1\right)}{2}
\end{align*}
where we have defined $\mu_{(1)}$ as

\begin{equation*}
\mu_{(1)} \equiv \min_{i \, \in  \, {1,2,...,N}} \left( \mu_{(i)} \right)
\end{equation*}
In addition (assuming that all values are exactly consecutive), their variance $V_{\mu}$ is given by

%/ Moeten we hier niet werken met delen door (N-1) trouwens ipv N?

\begin{align*}
V_{\mu} &= \frac{1}{N} \left( \sum_{j=1}^{N} \, \mu_{(j)}^2 \right) - M_{\mu}^2 \\
&= \frac{1}{N} \, \left( \sum_{j=0}^{N-1} \, \left( \mu_{(1)} + j \, \Delta \right)^2 \right) - M_{\mu}^2 \\
&= \frac{1}{N} \, \left( \sum_{j=0}^{N-1} \, \left( \mu_{(1)}^2 + 2 j \, \Delta \, \mu_{(1)} + j^2 \, \Delta^2 \right) \right) - M_{\mu}^2 \\
&= \mu_{(1)}^2 + \frac{2 \, \Delta \, \mu_{(1)}}{N} \, \left( \sum_{j=1}^{N-1} \, j \right) + \frac{\Delta^2}{N} \, \left( \sum_{j=1}^{N-1} \, j^2 \right) - M_{\mu}^2 \\
&= \mu_{(1)}^2 + \Delta \, \mu_{(1)} \, \left( N - 1 \right) + \frac{\Delta^2 \, \left( N-1 \right) \left( 2N-1\right)}{6}- M_{\mu}^2 \\
&= \frac{\Delta^2 \, \left( N-1 \right) \left( 2N-1\right)}{6} - \frac{\Delta^2 \left(N-1\right)^2}{4} \\
&= \frac{\Delta^2 \, (N-1)(N+1)}{12} \\
&\approx \frac{\Delta^2 \, N^2}{12}
\end{align*}
Hence, their standard deviation is $S_{\mu} \approx \frac{\Delta N}{2 \sqrt{3}}$.

Using the $N$ pairs, we assume as a linear interpolator in this example a linear regression model that links the simulated statistics to the true underlying $\mu$: $\bar{y}^{\mathrm{sim}}_j = \beta_0 + \beta_1 \mu_j + \epsilon_j$, with $\epsilon_j \sim N \left( 0, \frac{s^2}{T_{\mathrm{sim}}} \right)$. Obviously, $\beta_0 = 0$ and $\beta_1 = 1$. 

Given $\bar{y}$, $N$ selected prepaid points and the fitted linear regression model, we know from linear regression theory that:

\begin{align*}
\left( \begin{matrix} \hat{\beta}_0 \\ \hat{\beta}_1 \end{matrix} \right) &\sim N_2 \left( \left( \begin{matrix} 0 \\ 1 \end{matrix} \right), \left( \begin{matrix} \sigma_0^2 & \sigma_{01} \\ \sigma_{01} &  \sigma_1^2 \end{matrix} \right) \right),
\end{align*}
where 0 and 1 are the true $\beta_0$ and $\beta_1$ and

\begin{align*}
 \sigma_0^2 &= \mbox{Var}(\hat{\beta}_0|\bar{y}) \approx \frac{s^2}{T_{\mathrm{sim}}} \left( \frac{1}{N} + \frac{12 M_{\mu}^2}{\Delta^2 N^3} \right) \\
 \sigma_1^2 &= \mbox{Var}(\hat{\beta}_1|\bar{y}) \approx \frac{s^2}{T_{\mathrm{sim}}}  \frac{12}{\Delta^2 N^3} \\
 \sigma_{01} &= \mbox{Cov}(\hat{\beta}_0,\hat{\beta}_1|\bar{y}) = - M_{\mu} \sigma_1^2 \approx - \frac{s^2}{T_{\mathrm{sim}}}  \frac{12 M_{\mu}}{\Delta^2 N^3}.
\end{align*}
The distribution is assumed to hold for repeated simulations of the replicated statistics in the prepaid grid.

Because we work with linear regression, the optimization problem is simple. In this case, the optimal value of $\mu$ for a given $\bar{y}$ can be found by inverting the regression line: 
$$\hat{\mu} = \frac{\bar{y}-\hat{\beta}_0}{\hat{\beta}_1}.$$ 

Next, we can study the properties of $\hat{\mu}$. We begin by calculating the conditional mean $E(\hat{\mu}|\bar{y})$ and conditional variance $\mbox{Var}(\hat{\mu}|\bar{y})$. Hence, we treat the observed data (or sample average) as given and fixed. These expectations are taken over different simulations of $\bar{y}^{\mathrm{sim}}_j$'s in the prepaid grid. Before giving the expressions, it is useful to note that

\begin{align*}
\left( \begin{matrix} \bar{y} - \hat{\beta}_0 \\ \hat{\beta}_1 \end{matrix} \right) &\sim N_2 \left( \left( \begin{matrix} \bar{y} \\ 1 \end{matrix} \right), \left( \begin{matrix} \sigma_0^2 & -\sigma_{01} \\ -\sigma_{01} &  \sigma_1^2 \end{matrix} \right) \right).
\end{align*}
Now, using the approximations given in \cite{mood_introduction_1974} for ratios of random variables, we find that:

\begin{align*}
E(\hat{\mu}|\bar{y}) &= E\left( \frac{\bar{y}-\hat{\beta}_0}{\hat{\beta}_1} | \bar{y} \right) \\
&\approx \frac{E(\bar{y}-\hat{\beta}_0|\bar{y})}{E(\hat{\beta}_1|\bar{y})}-\frac{1}{E(\hat{\beta}_1|\bar{y})^2} \mbox{Cov}(\bar{y}-\hat{\beta}_0,\hat{\beta}_1|\bar{y})+\frac{E(\bar{y}-\hat{\beta}_0|\bar{y})}{E(\hat{\beta}_1|\bar{y})^3}\mbox{Var}(\hat{\beta}_1|\bar{y}) \\
&\approx \frac{\bar{y}}{1}-\frac{1}{1^2} \frac{s^2}{T_{\mathrm{sim}}}  \frac{12 M_{\mu}}{\Delta^2 N^3} + \frac{\bar{y}}{1^3} \frac{s^2}{T_{\mathrm{sim}}}  \frac{12}{\Delta^2 N^3} \\
&= \bar{y} \left(1 + \frac{s^2}{T_{\mathrm{sim}}}\frac{12}{\Delta^2 N^3} \right) -  \frac{M_{\mu}}{T_{\mathrm{sim}}}\frac{12 s^2}{\Delta^2 N^3}
\end{align*}
and

\begin{align*}
\mbox{Var}(\hat{\mu}|\bar{y}) &= \mbox{Var}\left( \frac{\bar{y}-\hat{\beta}_0}{\hat{\beta}_1} | \bar{y} \right) \\
& \approx \frac{E(\bar{y}-\hat{\beta}_0|\bar{y})^2}{E(\hat{\beta}_1|\bar{y})^2} \left( \frac{\mbox{Var}(\bar{y}-\hat{\beta}_0|\bar{y})}{E(\bar{y}-\hat{\beta}_0|\bar{y})^2} + \frac{\mbox{Var}(\hat{\beta}_1|\bar{y})}{E(\hat{\beta}_1|\bar{y})^2} - \frac{2 \mbox{Cov}(\bar{y}-\hat{\beta}_0,\hat{\beta}_1|\bar{y})}{E(\bar{y}-\hat{\beta}_0|\bar{y})E(\hat{\beta}_1|\bar{y})}\right) \\
&= \frac{\bar{y}^2}{1^2} \left( \frac{\sigma_0^2}{\bar{y}^2} + \frac{\sigma^2_1}{1^2} - \frac{2(-\sigma_{01})}{\bar{y} \cdot 1} \right) \\
&= \sigma_0^2 + \bar{y}^2 \sigma_1^2 - 2 \bar{y} M_{\mu} \sigma_1^2 \\
&\approx \frac{s^2}{T_{\mathrm{sim}} N} \left( 1 + \frac{12 M^2_{\mu}+12\bar{y}^2-24 \bar{y} M_{\mu}}{\Delta^2 N^2}  \right) \\
&= \frac{s^2}{T_{\mathrm{sim}} N} \left( 1 + \frac{12 \left( M_{\mu} - \bar{y} \right)^2}{\Delta^2 N^2}  \right).
\end{align*}
Invoking the double expectation theorem to arrive at the unconditional expectations, we have:

\begin{align}
E(\hat{\mu})& = E \left[ E(\hat{\mu}|\bar{y}) \right] \nonumber \\
&\approx E(\bar{y}) \left(1 + \frac{s^2}{T_{\mathrm{sim}}}\frac{12}{\Delta^2 N^3} \right) -  \frac{M_{\mu}}{T_{\mathrm{sim}}}\frac{12 s^2}{\Delta^2 N^3} \nonumber  \\
&= \mu \left(1 + \frac{s^2}{T_{\mathrm{sim}}}\frac{12}{\Delta^2 N^3} \right) -  \frac{M_{\mu}}{T_{\mathrm{sim}}}\frac{12 s^2}{\Delta^2 N^3} \nonumber  \\
&= \mu - \frac{\alpha}{T_{\mathrm{sim}}}\frac{12 s^2}{\Delta^2 N^3} \label{eq:mutoy},
\end{align}
where $\alpha = M_{\mu} - \mu$, that is, the difference between the true value and the mean of the selected nearest neighbors $\mu$'s.
Likewise, we can derive:

\begin{align}
\mbox{Var}(\hat{\mu}) &= E \left[ \mbox{Var}(\hat{\mu}|\bar{y}) \right] + \mbox{Var} \left[ E(\hat{\mu}|\bar{y})\right] \nonumber \\
&\approx \frac{s^2}{T_{\mathrm{sim}} N} \left( \frac{12(\frac{s^2}{T_{\mathrm{obs}}}+\mu^2)+12M_{\mu}^2-24M_{\mu}\mu}{\Delta^2 N^2}+1 \right) + \frac{s^2}{T_{\mathrm{obs}}}\left(1+\frac{12 s^{2}}{T_{\mathrm{sim}}\Delta^2 N^3} \right)^2 \nonumber \\
&= \frac{s^2}{T_{\mathrm{obs}}}+\frac{24s^4}{ T_{\mathrm{obs}} T_{\mathrm{sim}}\Delta^2 N^3}+\frac{144s^6}{T_{\mathrm{obs}}{T_\mathrm{sim}}^2 \Delta^4 N^6}+\frac{s^2}{T_{\mathrm{sim}}N}+ \frac{12s^2(\frac{s^2}{T_{\mathrm{obs}}}+\mu^2)+12 s^2 M_{\mu}^2-24 s^2 M_{\mu}\mu}{T_{\mathrm{sim}} \Delta^2 N^3} \nonumber \\
&= \frac{s^2}{T_{\mathrm{obs}}}+\frac{s^2}{T_{\mathrm{sim}}N}+\frac{24s^4}{ T_{\mathrm{sim}} T_{\mathrm{obs}} \Delta^2 N^3}+\frac{144s^6}{T_{\mathrm{sim}}^2 T_{\mathrm{obs}} \Delta^4 N^6}+ \frac{12s^2}{T_{\mathrm{sim}} \Delta^2 N^3} \left(  \frac{s^2}{T_{\mathrm{obs}}} + (\mu - M_{\mu} )^2 \right) \nonumber \\
&= \frac{s^2}{T_{\mathrm{obs}}}+\frac{s^2}{T_{\mathrm{sim}}N}+\frac{12s^2(\mu-M_{\mu})^2}{T_{\mathrm{sim}} \Delta^2 N^4}+\frac{36s^4}{ T_{\mathrm{sim}} T_{\mathrm{obs}} \Delta^2 N^3}+\frac{144s^6}{T_{\mathrm{sim}}^2 T_{\mathrm{obs}} \Delta^4 N^6} 
\label{eq:vartoy}
\end{align}

\begin{figure}[ht]
  \centering
    \includegraphics[width=1\textwidth]{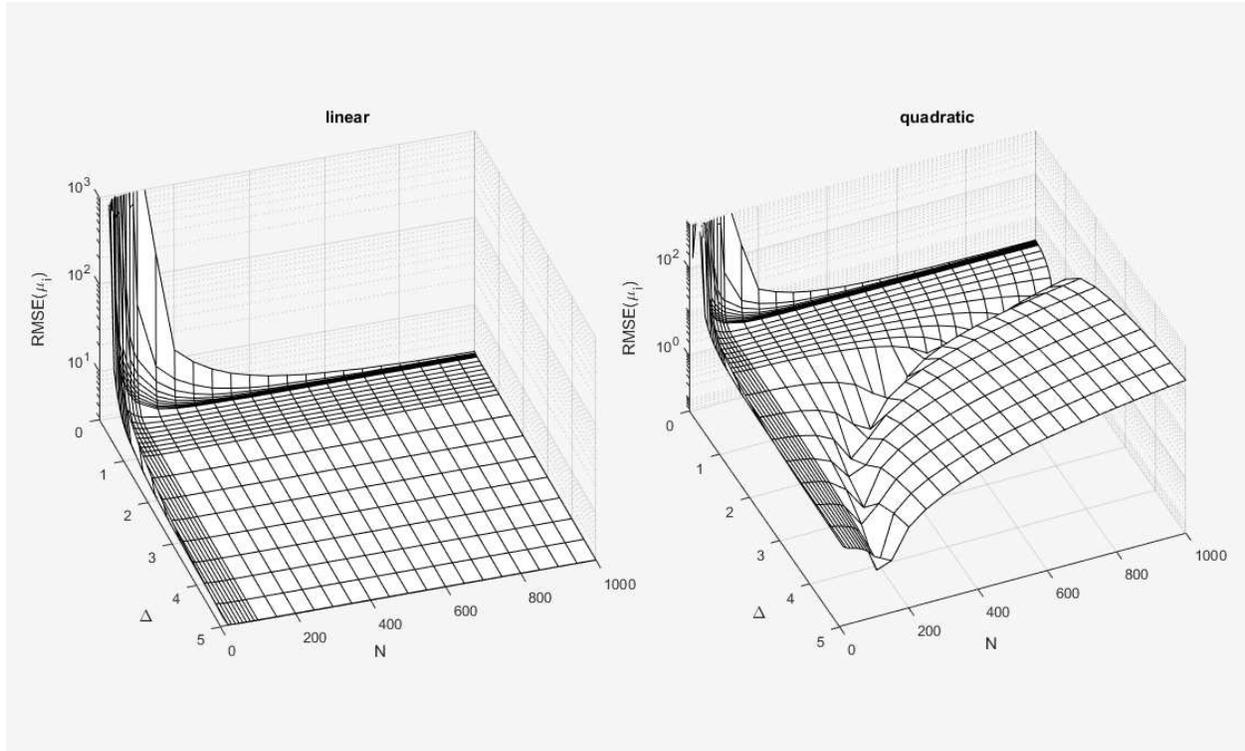}
     \caption{RMSE (based on a simulation study) of the toy example estimation as function of the gap size ($\Delta$) and number of nearest neighbors selected to carry out the interpolation ($N$). The left panel is called situation 1 in which $s^{\mathrm{obs}}= \bar{y}$ and the right panel is situation 2 ($s^{\mathrm{obs}}= \bar{y}^2$). For the second situation, the trade-off between $\Delta$ and $N$ is clearly visible.}
     \label{fig:toyexample}
\end{figure}

From Equation~\ref{eq:mutoy}, we learn that if there is no systematic deviation in the selection of $\mu$-grid points, the prepaid estimator is unbiased. In the other case, the is bias decreases with $T_{\mathrm{sim}}$ but is proportional to $s^2$. In Equation~\ref{eq:vartoy}, the leading term of the variance is $\frac{s^2}{T_{\mathrm{obs}}}$, which is the same as in classical estimation theory. For the other terms, they all have $T_{\mathrm{sim}}$ (or a power of it) in the denominator. Because  $T_{\mathrm{sim}}$ is usually quite large, these terms tend to be in general of lesser importance. However, some terms also have both $N$ (the number of selected nearest neighbor grid points) and $\Delta$ (the gap size). It is worthwhile to note that increasing the resolution (i.e., decreasing $\Delta$), while keeping $N$ constant, will increase the additional terms and thus add to the error. The reason for this is that the interpolation is defined on a too small grid, leading to uncertainty in the estimated regression. This effect is illustrated in the left panel of Figure~\ref{fig:toyexample} in which the root mean square error (RMSE) is shown for the estimation of $\mu$ for different values of $N$ and $\Delta$. The plot is constructed by means of a simulation study, but confirms our analytical results.

\paragraph{Situation 2: $s^{\mathrm{obs}}= \bar{y}^2$} 

In the second situation, we will again estimate $\mu$ (the unknown mean of a unit variance normal), but in this case $s^{\mathrm{obs}}= \bar{y}^2$ is used as a statistic. Thus, the relation between the simulated statistics $\bar{y}^{\mathrm{sim}^2}$ and $\mu$ is quadratic (and thus nonlinear). Again we use a local linear approximation. Clearly, this approximation will only be approximately valid if we do not choose the area of approximation too large. However, unlike in the first situation, we do expect an additional effect of the approximation error.

No analytical derivations were made for this case, but we conducted a similar simulation study as in situation 1. The results (in terms of RMSE) are shown in the right panel of Figure~\ref{fig:toyexample}. As can be seen, there is a clear optimality trade-off visible between $\Delta$ and $N$. This can be explained as follows: Fix $N$ and then consider the gap size $\Delta$. If $\Delta$ is too small, we get a similar phenomenon as in the left panel, that is a large RMSE. However, if we take $\Delta$ too large, then the approximation error will dominate (because the linear interpolation misfits the quadratic relation). The optimal point will be different for different $N$.

This toy example demonstrates the sound theoretical foundations of the prepaid method in well-behaved situations. However, the question is how well the method performs for real life examples. Three hard problems will be studied next.

\subsection{Application 1: The Ricker model} \label{subs:ricker}
The basic model equations of the Ricker model is given in Equation~\ref{eqRicker}.
\subsubsection*{Synthetic likelihood estimation}
For the synthetic likelihood estimation ($\text{SL}^{\text{Orig}}$), we made use of the \verb+synlik+ package \cite{fasiolo_introduction_2014}. The synthetic likelihood $l_s$ for a data set with summary statistics $\boldsymbol{s}\textsuperscript{obs}$ and a certain parameter vector $\boldsymbol{\theta}=(r,\sigma,\phi)$ is given by

\begin{equation} \label{SL}
l_s(\theta)=-\frac{1}{2}\left(\boldsymbol{s}\textsuperscript{obs}-\hat{\boldsymbol{\mu}}_{\boldsymbol{\theta}}\right)^T \hat{\boldsymbol{\Sigma}}_{\boldsymbol{\theta}}^{-1}
\left(\boldsymbol{s}\textsuperscript{obs}-\hat{\boldsymbol{\mu}}_{\boldsymbol{\theta}}\right) -\frac{1}{2}\log \left|\hat{\boldsymbol{\Sigma}}_{\boldsymbol{\theta}}\right|,
\end{equation}
where $\hat{\boldsymbol{\mu}}_{\boldsymbol{\theta}}$ and $\hat{\boldsymbol{\Sigma}}_{\boldsymbol{\theta}}$ are the estimated mean and covariance of the summary statistics when Equation~\ref{eqRicker} is simulated multiple times with parameter $\boldsymbol{\theta}$.

The statistics used by the synthetic likelihood function were the average population size, the number of zeros, the autocovariances up to lag 5, the coefficients of the quadratic linear autoregression of $y_t^{0.3}$ and the coefficients of the cubic regression of the ordered differences $y_t-y_{t-1}$ on the observed values. 

%% FT: check the numbers below
For each data set we used the synthetic likelihood Markov chain Monte Carlo (MCMC) method with 30000 iterations, a burn in of 3 time steps and 500 simulations to compute each $\hat{\boldsymbol{\mu}}_{\boldsymbol{\theta}}$ and $\hat{\boldsymbol{\Sigma}}_{\boldsymbol{\theta}}$ \cite{fasiolo_introduction_2014}. We used the following prior:

\begin{equation}
\begin{split}
r & \sim \mathcal{U} \left(1,90\right) \\
\sigma & \sim \mathcal{U}\left(0.05,0.7\right)\\ 
\phi & \sim \mathcal{U}\left(0,20\right). 
\end{split}
\label{rickPrior}
\end{equation}
The \verb+synlik+ package generates the MCMC chain on a logarithmic scale, we estimated the parameters as the exponential of the posterior mean. To ensure convergence, only the last half of the chain is used (the last 15000 iterations).
%% FT: kan je niet beter alle draws exp() doen en dan posterior mean berekenen?
%% MM: dit werkte beste, heb verschillende mogelijkheden geprobeerd

\subsubsection*{Creation of the prepaid grid}
For the prepaid estimation, we used the same summary statistics as for the traditional synthetic likelihood, except for two differences. First, the coefficients of the cubic regression of the ordered differences $y_t-y_{t-1}$ on the observed values could not be used, because the observed values are not available when creating the prepaid grid. Second, we changed the number of zeros to the percentage of zeros to make this statistic independent of $T_{\mathrm{obs}}$ (as this may change depending on the observation). 

We filled the prepaid grid with 100000 parameter sets using the priors of Equation \ref{rickPrior}. To cover this grid as evenly as possible (and avoiding too large gaps), the uniform distribution was approximated using Halton sequences \cite{matlab_version_2016, kocis_computational_1997}. For each parameter set in the prepaid grid, we simulated a time series of length $10^7$ and used the summary statistics of this long time series as $\hat{\boldsymbol{\mu}}_{\boldsymbol{\theta}}$. 

Each time series was then split into series of length $T_{prepaid}=$ 100,1000 and 10000 which were used to compute the covariance $\hat{\boldsymbol{\Sigma}}_{\boldsymbol{\theta,T_{prepraid}}}$ for the statistics computed on data of these lengths. This means, for example, that we had 100000 series of length 100 to compute the covariance matrix for a certain parameter set for time series of length 100. If we need to estimate parameters of a time series with $T_{\mathrm{obs}}$ not equal to one of the $T_{\mathrm{prepaid}}$ lengths, we use the covariance matrix created with time series of length $T_{prepaid}$ which is closest to $T_{\mathrm{obs}}$ in logarithmic scale and adapt the covariance matrix into

\begin{equation} \label{eq:covPP}
\hat{\boldsymbol{\Sigma}}_{\boldsymbol{\theta},T_{\mathrm{obs}}}=\frac{T_{\mathrm{prepaid}}}{T_{\mathrm{obs}}} \hat{\boldsymbol{\Sigma}}_{\boldsymbol{\theta},T_{\mathrm{prepaid}}}
\end{equation}
The creation of the prepaid grid took approximately one day on a 3.4GHz 20-core processor.

To allow the estimation for a bigger range of parameters for the online estimation at www.prepaidestimation.org we created a new and bigger prepaid grid using the following priors:

\begin{equation}
\begin{split}
\text{log}(r) & \sim \mathcal{U} \left(\text{log}(1),log(200)\right) \\
\sigma & \sim \mathcal{U}\left(0.05,0.7\right)\\ 
\text{log}(\phi) & \sim \mathcal{U}\left(-2,7\right). 
\end{split}
\label{rickPriorOnline}
\end{equation}

We filled to prepaid grid with 100000 parameter sets and used this prior for the real life data set.

\subsubsection*{Prepaid estimation}
Four variants of prepaid estimation were implemented for this example. All use the negative synthetic likelihood as distance ($d\left(\boldsymbol{s^{\mathrm{sim}}},\boldsymbol{s^{\mathrm{obs}}}\right)$ as defined in the main text and Figure \ref{fig:method}). First, we do a nearest neighbor estimation $\text{SL}^{\text{GRID}}_{\text{ML}}$, without using any interpolation between the grid points of the prepaid data set. We compute the synthetic likelihood of all the prepaid parameters for the summary statistics of the test data set. The parameter vector with the highest likelihood, the so-called nearest neighbor may already be a good estimation. For a low number of time points $T_{\mathrm{obs}}$, it is to be expected that the error on the parameter estimate is much larger than the gaps in the prepaid grid, and in such a case, the $\text{SL}^{\text{GRID}}_{\text{ML}}$ estimation approach suffices. 

Second, a more accurate estimation can be acquired by interpolating between the parameter values in the prepaid grid ($\text{SL}^{\text{SVM}}_{\text{ML}}$). Therefore, we learn the relation between the parameters and the summary statistics: $\hat{f}_{svm}: \boldsymbol{\theta} \mapsto \boldsymbol{s}$. However, we only learn this relation in the region of interest, that is the 100 nearest neighbors according to the synthetic likelihood. For each summary statistic, we create, on the fly, a separate least squares support vector machine (LS-SVM) \cite{suykens_least_2002} using the 100 nearest neighbors. This machine learning technique is chosen as it is a fast non-linear method which generalizes well. We limit the predictions to the possible range of the summary statistics (e.g., to prevent a percentage of zeros, one of the statistics, larger than 1). 

We then use the differential evolution global optimizer \cite{storn_differential_1997} to find the maximum of:

\begin{equation}
l_s^{\mathrm{PP}}(\theta)=-\frac{1}{2}\left(\boldsymbol{s}\textsuperscript{obs}-\hat{f}_{svm}\left(\boldsymbol{\theta}\right)\right)^T \hat{\boldsymbol{\Sigma}}_{\boldsymbol{\theta},T_{obs}}^{-1}
\left(\boldsymbol{s}\textsuperscript{obs}-\hat{f}_{svm}\left(\boldsymbol{\theta}\right)\right) -\frac{1}{2}\log \left|\hat{\boldsymbol{\Sigma}}_{\boldsymbol{\theta},T_{obs}}\right|,
\end{equation}
where $\hat{\boldsymbol{\Sigma}}_{\boldsymbol{\theta},T_{obs}}$ is the covariance matrix of the statistics of the nearest neighbor as defined in Equation~\ref{eq:covPP}. The superscript "PP" is used to denote that we use the prepaid version of synthetic likelihood, and not the traditional version as used by \cite{wood_statistical_2010} (see Equation~\ref{SL}). The optimization process is constrained and we use the minima and maxima for each parameter of the 100 nearest neighbors as effective bounds. 

The $\text{SL}^{\text{SVM}}_{\text{ML}}$ approach makes use of a non-linear black box interpolator. However, we may also consider using a much faster linear regression (see also the toy example in Section~\ref{subs:toy}). Therefore, we will also compare the $\text{SL}^{\text{SVM}}_{\text{ML}}$ (and $\text{SL}^{\text{Grid}}_{\text{ML}}$) approach to a third option where we predict the summary statistics using a linear regression (called the $\text{SL}^{\text{Lin}}_{\text{ML}}$ approach).

Third, we can easily implement a prior for the likelihood in Equation \ref{SL}. This leads to a posterior given by

\begin{equation}
p(\boldsymbol{\theta}|\boldsymbol{s}^{\mathrm{obs}}) \propto p(\boldsymbol{\theta}) l_s(\boldsymbol{\theta}).
\label{eqPriorRick}
\end{equation}

The parameters will be estimated as the maximum a posteriori (MAP), as comparison to maximum likelihood estimation which is a maximum a posteriori with a uniform prior. Here we will apply this extension to the nearest neighbor estimation: $\text{SL}^{\text{GRID}}_{\text{MAP}}$. 

Lastly we will show that our prepaid method can also be used to cover multiple experimental set ups. Each experimental set up involves multiple conditions and may have varying constraints on the parameters of the model over these conditions. If, for one experimental set up, the conditions $c$ are independent, the likelihood of the whole experiment is
\begin{equation}
l_{s, experiment}(\boldsymbol{\theta}^{1},\boldsymbol{\theta}^{2},...,\boldsymbol{\theta}^{C})=\prod_{c=1}^C{l_{s, c}(\boldsymbol{\theta}^{c})}
\end{equation}
where $l_{s, c}(\boldsymbol{\theta}^{c})$ is the synthetic likelihood for condition $c$. This is equivalent to estimating each parameter  set $\boldsymbol{\theta}^{c}$ individually  for each condition $c$. If the conditions are not independent and we assume that some parameters are the same over conditions we adapt the prior to mimic these assumptions. For example, if we assume that the first parameter $\theta_1$ is the same over all conditions $C$ we formulate this as
\begin{equation}
p(\theta_1^1,\theta_1^2,...,\theta_1^C)=\prod_{c=1}^{C} \mathcal{N}\left(\frac{\theta_1^c-\bar{\theta}_1}{\sigma_{prior}}\right)
\end{equation}
where $\mathcal{N}$ is the standard normal distribution and $\bar{\theta_1}$ is the average of all $\theta_1^c$. The smaller the tuning parameter $\sigma_{prior}$, the more all $\theta_1^c$ will be forced to be equal. If $\sigma_{prior}$ is too large the estimation will not take into account the interdependence between the conditions. However, if it is too small we run into trouble with the sparsity of the prepaid grid. In the limit, where $\sigma_{prior}$ goes to zero, one point is chosen from the prepaid grid leading to equal parameters not only for first parameter but also for the others. $\sigma_{prior}$ can be easily tuned by simulating the experimental set up for a certain prepaid grid.

To further illustrate, we will apply this method to the Ricker model, assuming two conditions across which $r$ and $\sigma$ stay the same such that this prior is given by
\begin{equation}
p(\boldsymbol{\theta^1},\boldsymbol{\theta^2})=\mathcal{N}\left(\frac{r_1-\bar{r}}{\sigma_{prior}}\right)\mathcal{N}\left(\frac{r_2-\bar{r}}{\sigma_{prior}}\right)\mathcal{N}\left(\frac{\sigma_1-\bar{\sigma}}{\sigma_{prior}}\right)\mathcal{N}\left(\frac{\sigma_2-\bar{\sigma}}{\sigma_{prior}}\right)
\label{exp_ls}
\end{equation}
We first use the nearest neighbor approach $\text{SL}^{\text{GRID}}_{\text{ML}}$ to find the 1000 nearest neighbors for condition one and two separately and then we refine the parameters using prior \ref{exp_ls}. As we assume $r$ and $\sigma$ to be constant over the conditions, we take $\bar{r}$ and $\bar{\sigma}$ as final estimates for these parameters in each condition.

\subsubsection*{Test set}
As a test set we first used 100 random parameters created with the prior of Equation \ref{rickPrior}. To avoid problems with the borders we deleted parameters that where within 1\% range of the bounds. We simulated data sets for $T_{\mathrm{obs}}=\lbrace 10^2, 5\cdot 10^2, 10^3,10^4,10^5\rbrace$. For each data set we estimated parameters using the nearest neighbor ($\text{SL}^{\text{Grid}}_{\text{ML}}$) and the $\text{SL}^{\text{SVM}}_{\text{ML}}$ approach. For $T_{\mathrm{obs}}=10^5$, we also estimated the parameters using the $\text{SL}^{\text{Lin}}_{\text{ML}}$ approach. Due to time constraints, we only estimated parameters for the data with $T_{\mathrm{obs}}\leq 10^3$ using the traditional synthetic likelihood approach.

Next we also created test data sets from different priors for $T_{\mathrm{obs}}=10^2$ . Prior $P_1$ from Equation \ref{rickPrior} can also be written as 
\begin{equation}
\begin{split}
\frac{r-1}{90-1} & \sim Beta \left(1,1\right) \\
\frac{\sigma-0.05}{0.7-0.05} & \sim Beta\left(1,1\right)\\ 
\frac{\phi}{20} & \sim Beta\left(1,1\right). 
\end{split}
\label{priorP1}
\end{equation}
where $Beta$ is a beta distribution with parameters $\alpha=1$ and $\beta=1$. Similarly, we created a test set from prior $P_2$

\begin{equation}
\begin{split}
\frac{r-1}{90-1} & \sim B \left(10,10\right) \\
\frac{\sigma-0.05}{0.7-0.05} & \sim B\left(10,10\right)\\ 
\frac{\phi}{20} & \sim B\left(10,10\right), 
\end{split}
\label{priorP2}
\end{equation}

and prior $P_3$
\begin{equation}
\begin{split}
\frac{r-1}{90-1} & \sim B \left(2,10\right) \\
\frac{\sigma-0.05}{0.7-0.05} & \sim B\left(10,2\right)\\ 
\frac{\phi}{20} & \sim B\left(2,10\right). 
\end{split}
\label{priorP3}
\end{equation}

We will test if $\text{SL}^{\text{GRID}}_{\text{MAP}}$ performs best when the correct prior is used in the estimation process. Last we also created a test set for $T_{\mathrm{obs}}=10^2$ for an experimental set up with two conditions where $r$ and $\sigma$ are equal over the conditions.

\subsubsection*{Results}
\begin{figure}[H]
  \centering
    \includegraphics[width=1\textwidth]{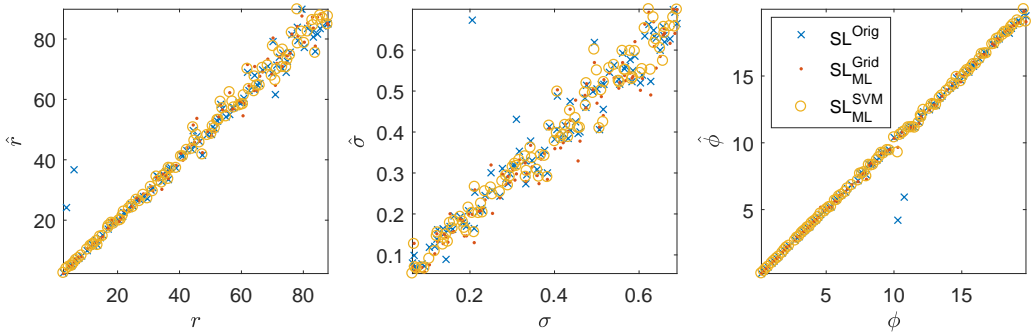}
     \caption{Estimated versus true parameters of the Ricker model of 100 data sets with $T_{\mathrm{obs}}=1000$. The $\text{SL}^{\text{Orig}}$ estimation has some problems with outliers.}
     \label{rickero}
\end{figure}

For the results, we will evaluate the methods on the following criteria: accuracy, speed, and coverage. 
\paragraph{Accuracy} To start off, we look at the recoveries for $T_{\mathrm{obs}}=10^3$ for all 100 simulated data sets and the three methods ($\text{SL}^{\text{Orig}}$,$\text{SL}^{\text{Grid}}_{\text{ML}}$ and $\text{SL}^{\text{SVM}}_{\text{ML}}$). Scatter plots are shown in Figure~\ref{rickero}. It can seen that the synthetic likelihood estimation leads to some clear outliers. One possible reason for the absence of outliers in the prepaid estimation is the fact that prepaid estimation from the start examines the whole grid and therefore has less problems with getting stuck in local optima.
%The only restriction that has to be taken into account is the resolution of the grid.
% FT: zou ik weglaten, want roept vragen op bij mij (welke resolutie, etc)

\begin{figure}[H]
  \centering
    \includegraphics[width=1\textwidth]{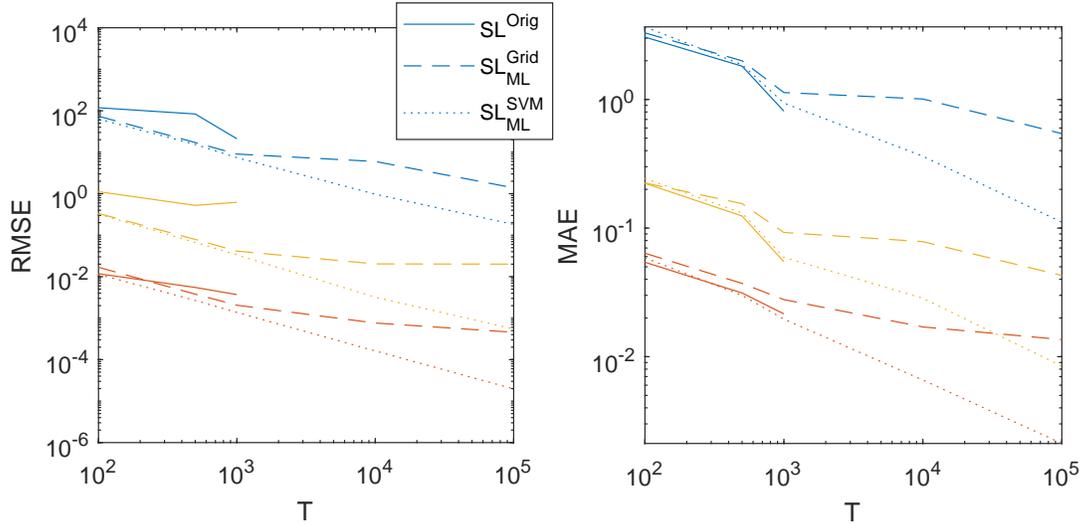}
     \caption{The accuracy of all estimation methods versus the number of time points $T_{\mathrm{obs}}$. The left panel shows the mean squared error, while the right panel shows the median absolute error. The three colors represent the three parameters. Blue lines refer to the parameter $r$, red lines to the parameter $\sigma$ and yellow lines to the parameter $\phi$. The solid line represents the original synthetic likelihood approach $\text{SL}^{\text{Orig}}$ (stopping at $T_{\mathrm{obs}}=10^3$), the dashed line the $\text{SL}^{\text{SVM}}_{\text{ML}}$ prepaid approach and the dotted line the $\text{SL}^{\text{SVM}}_{\text{ML}}$ prepaid approach.}
     \label{msetricker}
\end{figure}

%% FT: mean or median absolute error?
More generally, we plotted the accuracy of each of the methods as a function of time series length $T_{\mathrm{obs}}$ in Figure~\ref{msetricker}. The left panel shows the root mean square error (RMSE), while the right panel shows the median absolute error (MAE). We decided to look at the MAE because the few outliers for $\text{SL}^{\text{Orig}}$ (which were shown Figure~\ref{rickero}) may inflate the RMSE of the synthetic likelihood disproportionally, which happens to a certain extent. However, very similar conclusions can be drawn for both performance measures. In general, accuracy increases when $T_{\mathrm{obs}}$ increases (i.e., both RMSE and MAE decreases). For RMSE, our SVM prepaid method clearly outperforms the traditional synthetic likelihood method $\text{SL}^{\text{Orig}}$ for every $T_{\mathrm{obs}}$ and every parameter. For $T_{\mathrm{obs}}=\lbrace 5\cdot 10^2, 10^3\rbrace$, also the $\text{SL}^{\text{Grid}}_{\text{ML}}$ prepaid approach leads for every parameter to a lower RMSE compared to the synthetic likelihood. For all $T_{\mathrm{obs}}$, the $\text{SL}^{\text{SVM}}_{\text{ML}}$ prepaid leads to a higher accuracy compared to the $\text{SL}^{\text{Grid}}_{\text{ML}}$ prepaid and this difference becomes larger for a larger $T_{\mathrm{obs}}$. For MAE, the $\text{SL}^{\text{SVM}}_{\text{ML}}$ prepaid method and the original synthetic likelihood $\text{SL}^{\text{Orig}}$ show a very similar accuracy (for $T_{\mathrm{obs}} \leq 10^3$). Both outperform the $\text{SL}^{\text{Grid}}_{\text{ML}}$ prepaid.

%%FT: lijnen duidelijker; T_obs op x-as

\begin{figure}[H]
  \centering
    \includegraphics[width=1\textwidth]{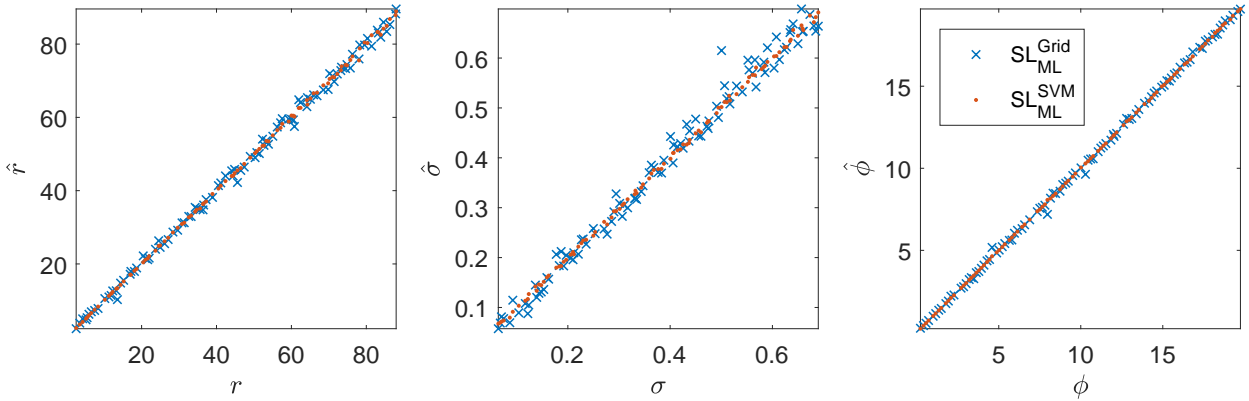}
     \caption{The estimation of the three parameters of the Ricker model of 100 data sets with $T_{\mathrm{obs}}=10^5$. The $\text{SL}^{\text{SVM}}_{\text{ML}}$ estimation clearly outperforms the $\text{SL}^{\text{Grid}}_{\text{ML}}$ estimation. }
     \label{rickers}
\end{figure}

The largest attainable accuracy for the $\text{SL}^{\text{Grid}}_{\text{ML}}$ prepaid approach is limited by the spacing of the prepaid grid. If we had created an equally spaced grid of $T_{\mathrm{obs}}=10^5$ points using the prior in Equation~\ref{rickPrior}, we would have the following gaps in each of the three parameter dimensions: 

\begin{equation}
\begin{split}
\Delta_r & = \frac{90-1}{(10^5)^{1/3}} =1.9\\
\Delta_{\sigma} & =\frac{0.7-0.05}{(10^5)^{1/3}} =0.01\\
\Delta_{\phi} & =\frac{20-0}{(10^5)^{1/3}} = 0.4.
\end{split}
\label{rickD}
\end{equation}
We do not have an equally spaced grid, but it is expected that the quasi Monte Carlo distribution of points creates expected gaps close to the ones in Equation~\ref{rickD}. Therefore, it is no coincidence that the best possible RMSE using the $\text{SL}^{\text{Grid}}_{\text{ML}}$ prepaid approach has the same order of magnitude as the gap size $\Delta$, as can be seen in Table~\ref{tabr2} for the case of $T_{\mathrm{obs}}=10^5$. However, Table~\ref{tabr2} also show that the $\text{SL}^{\text{SVM}}_{\text{ML}}$ prepaid approach leads to a much lower RMSE. The difference between the $\text{SL}^{\text{Grid}}_{\text{ML}}$ and the $\text{SL}^{\text{SVM}}_{\text{ML}}$ prepaid approach for $T_{\mathrm{obs}}=10^5$ is further visualized in Figure~\ref{rickers}. 

The results in Table~\ref{tabr2} also show the need for a non-linear interpolator for the prepaid method. The RMSE of a linear regression interpolator ($\text{SL}^{\text{Lin}}_{\text{ML}}$) is much larger than that of the SVM prepaid. 

\begin{table}
\caption{RMSE for the estimation of the parameters of the Ricker model for $T=10^5$ using the $\text{SL}^{\text{Grid}}_{\text{ML}}$, $\text{SL}^{\text{SVM}}_{\text{ML}}$ and $\text{SL}^{\text{Lin}}_{\text{ML}}$ prepaid methods.}
\begin{center}
\begin{tabular}{ l r r r}
\hline   & r & $\sigma$ & $\phi$  \\ \hline
  $\text{SL}^{\text{Grid}}_{\text{ML}}$ & 1.2 & 0.021 & 0.14\\
  $\text{SL}^{\text{SVM}}_{\text{ML}}$  & 0.43 & 0.0044 & 0.023 \\
  $\text{SL}^{\text{Lin}}_{\text{ML}}$  & 0.54 & 0.013 & 0.091 \\ \hline
  \end{tabular}
\end{center}
\label{tabr2}
\end{table}

In sum, we can conclude that the prepaid estimation methods lead to better, or at least similar, results as the traditional synthetic likelihood. 

\paragraph{Speed} The largest improvement of the prepaid method over synthetic likelihood  is in computational speed: The prepaid method is many times faster than synthetic likelihood. Consider Figure~\ref{rickert} in the main text where it is shown that the $\text{SL}^{\text{Grid}}_{\text{ML}}$ prepaid method is finished before a single iteration of the 30000 iterations are done by the $\text{SL}^{\text{Orig}}$ method. While the $\text{SL}^{\text{Grid}}_{\text{ML}}$ and the $\text{SL}^{\text{SVM}}_{\text{ML}}$ prepaid methods are finished in respectively 0.044 and 3.7 seconds, independent of the time series length $T_{\mathrm{obs}}$, the $\text{SL}^{\text{Orig}}$ method grows slower with an order of magnitude of $T_{\mathrm{obs}}$. In each $\text{SL}^{\text{Orig}}$ iteration one needs to simulate multiple time series with length $T_{\mathrm{obs}}$. The larger $T_{\mathrm{obs}}$, the slower the estimation. While the synthetic likelihood needs approximately one and a half hour to estimate the parameters for a time series with length $T_{\mathrm{obs}}=10^3$. The  $\text{SL}^{\text{Grid}}_{\text{ML}}$ prepaid estimation still finishes in 0.044 s, which is more than $10^5$ times faster. The speed up factors are presented in Table~\ref{tabr} and as can be seen from Figure~\ref{msetricker}, there is not loss of accuracy. The speed up would reach millions, if we had the time to run the synthetic likelihood method for longer time series.

\paragraph{Coverage} Next, we look at the coverage rates of the $95\%$ confidence intervals as obtained with the bootstrap in combination with the  prepaid method. To estimate a $95\%$ confidence interval of the estimates for the prepaid method, a parametric bootstrap with $B=1000$ bootstrap samples was used. 

For the prepaid version the estimate for the observed data set was obtained using the  $\text{SL}^{\text{SVM}}_{\text{ML}}$ approach and the bootstrap estimates were commonly obtained using the  $\text{SL}^{\text{Grid}}_{\text{ML}}$ prepaid method applied to the bootstrap data sets. However, if in the first 100 bootstraps only half of the nearest neighbors where unique points, the bootstrap distribution could be considered questionable. This behavior is to be expected for larger sample sizes $T_{\mathrm{obs}}$, because the true bootstrap distribution is very peaked so that every bootstrap sample will have the same nearest neighbor grid point. When this occurs, we would estimate the parameters of each bootstrap using differential evolution, using the SVM created by the original 100 nearest neighbors.
%% vorige paragraaf moeten we nog eens bespreken
%% het gaat ook om tijdreeksen, hoe wordt bootstrap hier toegepast
%% ook belangrijk bij parametric bootstrap?

Alternatively, for the synthetic likelihood approach (using MCMC) we computed the $95\%$ confidence interval by calculating the 0.025 and 0.975 quantiles of the last half of the posterior samples. 

The coverage results for the test set of 100 parameters are shown for three different values of $T_{\mathrm{obs}}$ in Table~\ref{tabrc}. It can be seen that for both methods, the coverage is close to the nominal level of $95\%$, but the coverage of the prepaid method is slightly better.

\paragraph{Prior}
In this paragraph we show how we can benefit from using the correct prior. We estimate the parameters of the three testsets for $T_{\mathrm{obs}}=100$, created with uniform prior $P_1$ from Equation \ref{eqPriorRick} and beta distribution priors $P_2$ and $P_3$ from Equations \ref{priorP2} and \ref{priorP3}. We estimated all three data sets using maximum a posteriori estimation $\text{SL}^{\text{GRID}}_{\text{MAP}}$ using all three priors. The results are shown in Table
\ref{tablePrior1}. Using the correct prior leads, as expected, to the best results.

\paragraph{Parameter constraints across conditions}
We estimated the parameters for a two condition experimental set up with equal $r$ and $\sigma$, with and without the prior from Equation \ref{exp_ls} (parameter $\sigma_{prior}$ was tuned on 100 similar simulated data sets). The results are shown in Table \ref{tableExp1}. Using the prior from Equation \ref{exp_ls}, which implements the parameter constraints of the experimental set up, leads, as expected, to better results for each parameter. Even for $\phi$, which is absent in the prior, we find better results.

\paragraph{Real life data set}
The results for the estimation of the population dynamics of the  Chilo partellus \cite{yonow_potential_2017,taneja_methods_1985}, using the prior from Equation \ref{rickPriorOnline} can be found in Table \ref{tabReal}. For the prepaid, we estimated the parameters using the methods  online at www.prepaidestimation.org. All estimations are similar and have overlapping confidence intervals. The prepaid estimation is however significantly faster.

\begin{table}
\caption{Average time in seconds needed for the  $\text{SL}^{\text{Orig}}$ estimation for multiple $T_{\mathrm{obs}}$ and the speed up for the  $\text{SL}^{\text{Grid}}_{\text{ML}}$ and  $\text{SL}^{\text{SVM}}_{\text{ML}}$ methods. The time for $T_{\mathrm{obs}}=10^4$ and $T_{\mathrm{obs}}=10^5$ was not measured, so these values are estimated and between brackets. (Figure~\ref{msetricker} shows the corresponding accuracies.)}
\begin{center}
\begin{tabular}{ l r r r r r}
\hline  $T_{\mathrm{obs}}$ & $10^2$ & $5\cdot 10^2$ & $10^3$ & $10^4$ & $10^5$  \\ \hline
  time $\text{SL}^{\text{Orig}}$  & 716 s & 3549 s & 5841 s & (50000 s) & (500000 s) \\
  $\text{SL}^{\text{GRID}}_{\text{ML}}$ times faster  & 16273 & 80659 & 132750 & (1000000) & (10000000) \\
  $\text{SL}^{\text{SVM}}_{\text{ML}}$ times faster &  194     &   959    & 1578   & (10000) & (100000) \\ \hline
\end{tabular}
\end{center}
\label{tabr}
\end{table}

\begin{table}
\caption{The effective coverages of the test set for different $T_{\mathrm{obs}}$.}
\begin{center}
\begin{tabular}{l r r r r}
\hline  &$T_{\mathrm{obs}}$ & r & $\sigma$ & $\phi$  \\ \hline
  &$10^2$ & 0.9 & 0.89  & 0.93  \\
   $\text{SL}^{\text{Orig}}$ &$5\cdot 10^2$ & 0.94 & 0.92 & 0.94 \\
  & $10^3$ & 0.92 & 0.91 & 0.92 \\ \hline
  &   $10^2$ & 0.95 & 0.84  & 0.97  \\
  prepaid&$5\cdot 10^2$ & 0.96 & 0.94 & 0.96 \\
  & $10^3$ & 0.97 & 0.95 & 0.97 \\ \hline
\end{tabular}
\end{center}
\label{tabrc}
\end{table}

\begin{table}
\caption{Population dynamics of the Chilo partellus \cite{yonow_potential_2017,taneja_methods_1985}. We show the estimates, the 95\% confidence intervals and computation time of the prepaid and synthetic likelihood estimation techniques.}
\begin{center}
\begin{tabular}{l r r r r}
\hline  & r & $\sigma$ & $\phi$ & Time (in seconds)  \\ \hline
   $\text{SL}^{\text{Orig}}$ &1.05 (1.01-- 1.1) & 0.41  (0.31 -- 0.51)& 248.17 (139.53 -- 493.2)& 830 \\
  $\text{SL}^{\text{GRID}}_{\text{ML}}$ & 1.10 (1.06-- 1.34) & 0.43  (0.30 -- 0.54)& 140.60 (43.94 -- 208.19)& 0.2 \\ 
   $\text{SL}^{\text{SVM}}_{\text{ML}}$&   1.06 (1.01-- 1.24) & 0.41  (0.21 -- 0.56)& 176.15 (19.27 -- 427.65)& 4 \\ \hline
\end{tabular}
\end{center}
\label{tabReal}
\end{table}

\begin{table}
\caption{RMSE of $\text{SL}^{\text{GRID}}_{\text{MAP}}$ estimation of test sets with $T_{\mathrm{obs}}=100$ created with priors $P_1$, $P_2$ and $P_3$ and estimated by using priors $P_1$, $P_2$ and $P_3$. For each test set and parameter the best result is shown in bold.}
\begin{center}
\begin{tabular}{ l r r r r r r r r r}
\hline   &\multicolumn{3}{c}{estimated with $P_1$}&\multicolumn{3}{c}{estimated with $P_2$} &\multicolumn{3}{c}{estimated with $P_3$}  \\
 parameter & r & $\sigma$ & $\phi$& r & $\sigma$ & $\phi$& r & $\sigma$ & $\phi$ \\ \hline
   test set created with $P_1$ &\textbf{ 8.2}  &  \textbf{ 0.13}   &  \textbf{0.53 }&  10 &    0.12 &    0.82&    16  &   0.17  &   0.94 \\
 test set created with $P_1$ &10&  0.13  &   0.55  &  \textbf{6.5}  &  \textbf{0.072 }&    \textbf{0.43}&    11   &  0.12  &  0.60 \\
test set created with $P_1$&  4.4 &   0.15&    0.33 &   6.9  &  0.19&     0.51 &    \textbf{3.5} &  \textbf{  0.065 } &  \textbf{ 0.28}\\ \hline
\end{tabular}
\end{center}
\label{tablePrior1}
\end{table}

\begin{table}
\caption{RMSE for Ricker model data where $T_{\mathrm{obs}}=100$ for an experimental set up with two conditions where $r$ and $\sigma$ are equal over the conditions. Parameters are estimated by using $\text{SL}^{\text{GRID}}_{\text{MAP}}$  with a flat prior (same as $\text{SL}^{\text{GRID}}_{\text{ML}}$)and with a prior from Equation \ref{exp_ls}}
\begin{center}
\begin{tabular}{ l r r r}
\hline   
 prior & r & $\sigma$ & $\phi$ \\ \hline
   flat prior & 88 &  0.17  & 0.42 \\
prior Equation \ref{exp_ls} &61&  0.11  &   0.36  \\ \hline
\end{tabular}
\end{center}
\label{tableExp1}
\end{table}

\subsection{Application 2: A stochastic model of community dynamics} \label{subs:comdyn}

%FT: er stond nog in de titel " TODO verwijzen naar prepaid achtige in zijn paper"
A second model we will apply our prepaid modeling technique to, is a stochastic dispersal-limited trait-based model of community dynamics \cite{jabot_stochastic_2010}. The data that will be modeled, are the abundances of species (hence a vector of frequencies, in which each component is a different species). Each species in the local environment is assumed to have a competitive value dependent on its trait $u$, given by the filtering function

\begin{equation} \label{eq:filtering}
F(u)=1+A e^{-\frac{(u-h)^2}{2\sigma^2}}.
\end{equation}
Here $A$ is the maximal competitive advantage, $h$ is the optimal trait value in the local environment and $\sigma$ describes the width of the filtering function. At each time step, one individual from the local community dies. It is then replaced with a probability $1-\frac{I}{I+J+1}$ by a random descendant from the local pool. Here, $J$ is the size of the local community and $I$ is the fourth parameter to estimate, related to the amount of immigration from the regional pool into the local community. The probability that this descendant comes from a certain individual in the local community, is proportional to the competitiveness of this individual. With a probability of $\frac{I}{I+J+1}$, the dead individual is replaced by an immigrant from the regional pool. The distribution of traits $u$ of the individuals in the regional pool is assumed to be uniform over $u$. It is noteworthy that Jabot saw the necessity of reusing ABC simulations to reduce computation time in his recovery study \cite{jabot_stochastic_2010}. 

The model was simulated using the C++ code from the Easy ABC package \cite{jabot_easyabc:_2015} where a regional pool of $S=1000$ species was defined evenly spaced on the trait axis (i.e., the resolution) and $J=500$ was the size of the local community. 

\subsubsection*{ABC estimation}
We compare our prepaid method estimation with the Easy ABC package ($\text{ABC}^{\text{Orig}}$) \cite{jabot_easyabc:_2013, jabot_easyabc:_2015}. Because we work in a Bayesian framework, we first have to specify priors. As in Jabot et al. we use the following priors \cite{jabot_easyabc:_2015}:

\begin{equation}
\begin{split}
 \log (I)  & \sim \mathcal{U} \left(3,5\right) \\
\log(A) & \sim \mathcal{U}\left(\log(0.1),\log(5)\right) \\
h & \sim \mathcal{U}\left(-25,125\right)\\ 
\log(\sigma) & \sim \mathcal{U}\left(\log(0.5),\log(25)\right).  
\end{split}
\label{traitPrior}
\end{equation}
In this application, the parameter vector $\boldsymbol{\theta}$ is defined as follows: $\boldsymbol{\theta}=(\log(I),\log(A),h,\log(\sigma))$. 
To get the ABC algorithm to work, we compute four summary statistics: the richness of the community (number of living species), Shannon's index which measures the entropy of the community, and the mean and the skewness of the trait distribution of the community.
% kunnen we definitie van deze statistieken geven of zeggen waar definitie kan gevonden worden? doen ze zelf ook niet. heb richness wat uitgelegd, de rest is redelijk logisch?

The ABC algorithm we use applies a sequential parameter sampling scheme \cite{beaumont_adaptive_2009}. The sequence of tolerance bounds is given by $\rho=\{8,5,3,1,0.5,0.2,0.1\}$ and the algorithm proceeds to the next tolerance after 200 simulations which lead to summary statistics within the bounds. The last 200 simulations within the bounds represent the posterior, and the estimate of the parameter is given by the posterior mean.

\subsubsection*{Creation of the prepaid grid}
For the prepaid estimation, we used exactly the same summary statistics as the Easy ABC package. We filled the prepaid grid with $500,000$ parameter vectors using the priors of Equation~\ref{traitPrior}, but for most examples we will use a grid with only $100,000$ parameter vectors. To cover this grid as evenly as possible, the uniform distribution was approximated using Halton sequences \cite{matlab_version_2016, kocis_computational_1997} (in order to avoid gaps that may appear when Monte Carlo samples are used). The creation of the prepaid grid with $100,000$ parameter vectors took approximately 3 days on a 3.4GHz 20-core processor.

For the community dynamics models from Equations~\ref{eq:filtering} and~\ref{traitPrior}, there are several ways to simulate an almost infinitely large data set to achieve stable summary statistics. The first way is to increase the number of species $S$ and the size of the local pool $J$. Unfortunately some summary statistics (the richness and the entropy) are in some unknown way dependent on these parameters. As a result, the summary statistics of a simulation with $J=5000$ cannot be used to estimate the parameters for a setting where $J=500$. Therefore, we chose to fix the size of the local pool $J$ and the number of species $S$. It is very well possible that there are summary statistics which do not have this problem, making the prepaid grid much more universal. We chose however, for the sake of comparison with the easy ABC package to keep using these parameters.

A second way to simulate data with a very large sample size is by increasing the number of time steps. By estimating the summary statistics after each time step, when one individual from the local community dies and is replaced by another individual, we create a time series of summary statistics. Averaging the summary statistics over a sufficient large number of time points will lead to stable average values of these summary statistics. In our simulations, we applied some tinning by calculating the summary statistics every time after 500 species have died (the size of the community). The reasons is that there is not enough of variation in the summary statistics computed after the death of a single species. Next, we created time series of length $T=100,000$ ($5\cdot 10^7$ species will have been replaced) for the prepaid grid and used the average of these summary statistics as $\hat{\boldsymbol{\mu}}_{\boldsymbol{\theta}}$. Using this time series we also computed $\hat{\boldsymbol{\Sigma}}_{\boldsymbol{\theta},T_{\mathrm{prepaid}}}$ for $T_{\mathrm{prepaid}}=\{1,10,1000,10000\}$. $T_{\mathrm{prepaid}}=1$ is of course the setting for which the original trait model is described and for which the Easy ABC algorithm is tested. Additionally we also saved 1000 samples of time series of length $T_{\mathrm{prepaid}}=\{1,10,1000,10000\}$.

\subsubsection*{Prepaid estimation}
Contrary to the first application (the Ricker model), where we used a frequentist approach, for this community dynamics model we will follow a Bayesian approach. In Bayesian statistics, the focus is on the posterior distribution of the parameters $p(\boldsymbol{\theta}|\text{data})$, which is defined as follows:

\begin{equation}
p(\boldsymbol{\theta}|\text{data})\propto p(\text{data}|\boldsymbol{\theta}) \times p(\boldsymbol{\theta}),
\end{equation} 
where $p(\text{data}|\boldsymbol{\theta})$ is the likelihood and $p(\boldsymbol{\theta})$ the prior. As the likelihood, we will use the synthetic likelihood $p(\text{data}|\boldsymbol{\theta}) \approx L_s(\boldsymbol{\theta})=\exp(l_s(\boldsymbol{\theta}))$, where $l_s(\boldsymbol{\theta})$ is the synthetic log-likelihood as defined in Equation~\ref{SL} (based on the vector of summary statistics $\boldsymbol{s}^{\mathrm{obs}}$). Because we compress the data into summary statistics, the posterior we work with is actually an approximation to the true posterior: $p(\boldsymbol{\theta}|\boldsymbol{s}^{\mathrm{obs}})\approx p(\boldsymbol{\theta}|\text{data})$ (in case the summary statistics are sufficient statistics for $\boldsymbol{\theta}$, the approximation sign becomes an equality sign). The distributions from Equation~\ref{traitPrior} are the priors for the parameters. 

We have studied three variants of a Bayesian version of the prepaid method. These three versions will be discussed here in increasing order of complexity. We will denote the three variants as follows: $\text{SL}^{\text{Grid}}_{\text{PM}}$, $\text{ABC}^{\text{Grid}}_{\text{PM}}$, and $\text{ABC}^{\text{SVM}}_{\text{PM}}$. 

\paragraph{$\text{SL}^{\text{Grid}}_{\text{PM}}$} Because the priors are all uniform (and our prepaid grid is distributed following this prior), the posterior for a data set with summary statistic $\boldsymbol{s}$ at parameter $\boldsymbol{\theta}_p$ of the prepaid grid is proportional to

\begin{equation}
p(\boldsymbol{\theta}|\boldsymbol{s}^{\mathrm{obs}}) \propto L_s^{\mathrm{PP}}(\boldsymbol{\theta}),
\label{eqPSL}
\end{equation}
where $L_s^{\mathrm{PP}}(\boldsymbol{\theta})$ is the prepaid synthetic likelihood (i.e., with the mean statistics computed for a very large sample and a approximate covariance matrix given by Equation \ref{eq:covPP}). The posterior mean (PM) using prepaid synthetic likelihood can be estimated as:

\begin{equation}
\hat{\boldsymbol{\theta}}|\boldsymbol{s}^{\mathrm{obs}}=\frac{\sum_{p}{L_s^{\mathrm{PP}}(\boldsymbol{\theta}_p) \times \boldsymbol{\theta}_p}}
{\sum_{p}{L_s^{\mathrm{PP}}(\boldsymbol{\theta}_p)}}.
\end{equation}

\begin{figure}[ht]
  \centering
    \includegraphics[width=1\textwidth]{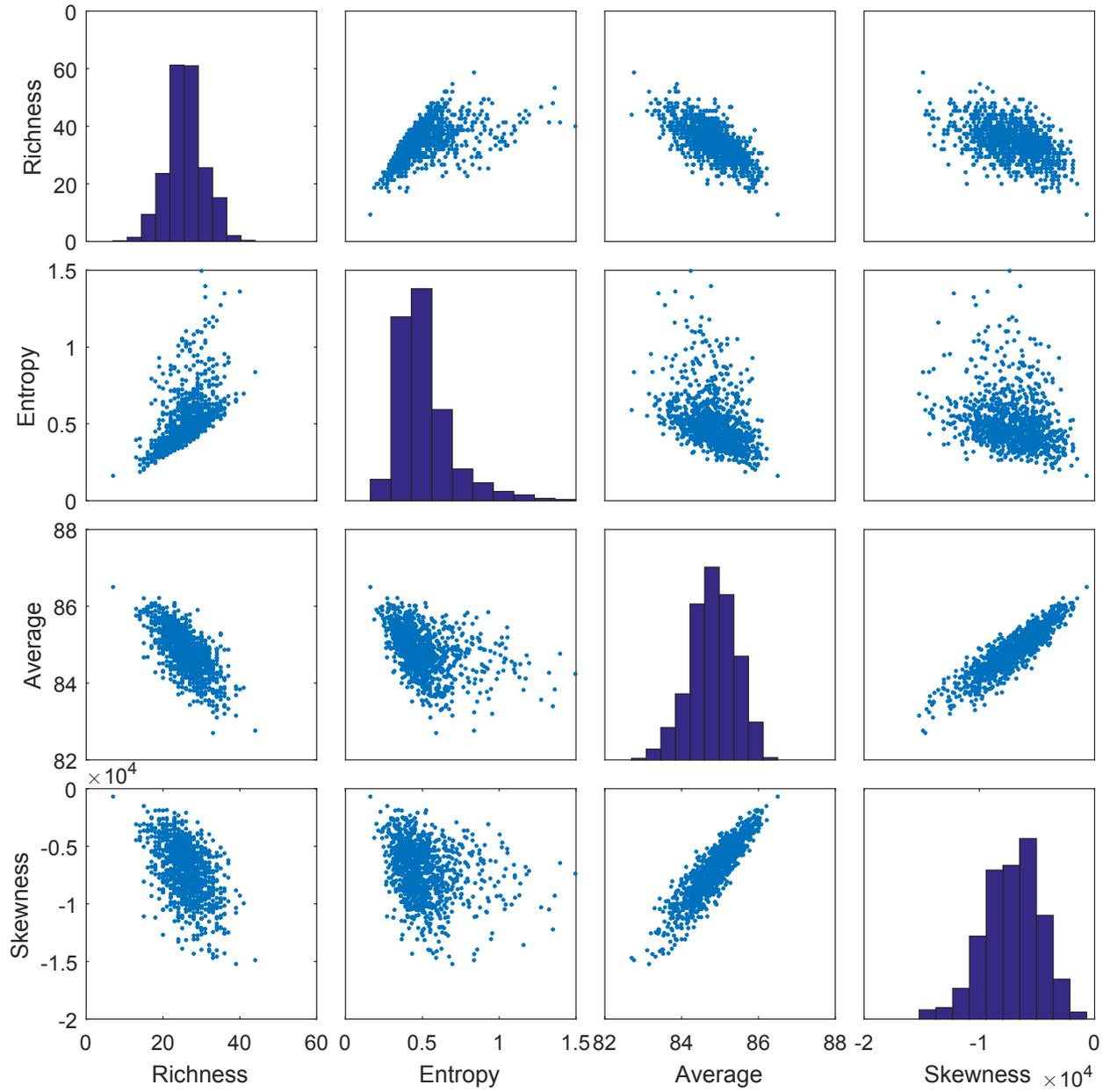}
     \caption{Samples for $T_{\mathrm{obs}}=1$ of the summary statistics of the trait model for parameter set
     $\log (I)=3.0621$, $\log(A)=0.8302$, $h=86.8924$ and $\log(\sigma)=-0.6899$.}
     \label{samplesNonNormal}
\end{figure}

\paragraph{$\text{ABC}^{\text{Grid}}_{\text{PM}}$} The prepaid synthetic likelihood approach works best if the assumption of normally distributed summary statistics is not too far off. However, as can be seen in Figure~\ref{samplesNonNormal}, this is not always the case for the trait model defined in Equation~\ref{eq:filtering}. Therefore, as an alternative procedure, we propose an Approximate Bayesian Computation (ABC) approach. First, we select a subset of nearest neighbors $\cal{S}$ from the prepaid set, such that for every $\boldsymbol{\theta}_q \in \cal{S}$, the synthetic likelihood value $L_s(\boldsymbol{\theta}_q)$ is highest and so that 

\begin{equation}
\frac{\sum_{q}{L_s^{\mathrm{PP}}(\boldsymbol{\theta}_q)}} {\sum_{p}{L_s^{\mathrm{PP}}(\boldsymbol{\theta}_p)}}<0.999,
\end{equation}
where the sum in the denominator runs across all grid points. In a sense, these are all the prepaid points in the $99.9\%$ expected coverage according to the posterior of Equation~\ref{eqPSL}. We denote the cardinality of $\cal{S}$ as $Q$. 

In a next step, we basically perform ABC with all the grid points belonging to the selected subset $\cal{S}$. However, there is an important issue we cannot overlook. When doing ABC, for a given parameter vector new data are simulated of the same size as the observed data. Unfortunately, our prepaid grid has correspondingly only very large data sets. To rectify this problem, so that ABC can applied without problems, we simulated during the construction of the prepaid grid, a set of $M=1000$ prepaid samples for several designated sample sizes (i.e., $T_{\mathrm{prepaid}}=\{1,10,1000,10000\}$). Let us denote with $\boldsymbol{s}_{q,i,T_{\mathrm{prepaid}}}$ the vector of statistics for prepaid grid point $q$, the $i$th simulation (with $i=1,\dots,M$) and sample size $T_{\mathrm{prepaid}}$.

Now, we can apply ABC to arrive at the posterior for $\boldsymbol{\theta}$; the method will be denoted as $\text{ABC}^{\text{Grid}}_{\text{PM}}$. For now we will assume that $T_{obs}$ is equal to one of the $T_{\mathrm{prepaid}}$ lenghts. We select the 1000 samples from this $Q \times 1000$ samples set that have the smallest Mahalonobis distance to the observed set of statistics $\boldsymbol{s}^{\mathrm{obs}}$:

\begin{equation}
\epsilon_{q,i,T_{\mathrm{prepaid}}}=(\boldsymbol{s}_{q,i,T_{\mathrm{ABC}}}-\boldsymbol{s}^{\mathrm{obs}})\boldsymbol{W}_{Q}^{-1}(\boldsymbol{s}_{q,i,T_{\mathrm{ABC}}}-\boldsymbol{s}^{\mathrm{obs}})
\label{epsTrait}
\end{equation}
here $\boldsymbol{W}_Q$ is given by the covariance over all grid points in $\cal{S}$ and over all 1000 replications (thus, $Q \times 1000$). The finally selected 1000 samples are then considered as a sample from the posterior. Note that the $\text{ABC}^{\text{Grid}}_{\text{PM}}$ method does not require us to progressively strengthen the tolerances, as in traditional $\text{ABC}^{\text{Orig}}$ (governed by the tolerance parameter $\rho$). If the observed sample size $T_{obs}$ is not equal to one of the $T_{\mathrm{prepaid}}$ lengths, then one can use the samples for length $T_{\mathrm{prepaid}}$ which is closest to $T_{obs}$ in logaritmic scale and later adjust the posterior samples such that the posterior mean stays the same, but the posterior covariance matrix changes to

\begin{equation} \label{postevt}
\hat{\boldsymbol{\Sigma}}_{posterior,T_{\mathrm{obs}}}=\frac{T_{\mathrm{prepaid}}}{T_{\mathrm{obs}}} \hat{\boldsymbol{\Sigma}}_{posterior,T_{\mathrm{prepaid}}}
\end{equation}
We advise to save samples for enough different $T_{\mathrm{prepaid}}$ such that this correction is only marginal.

\paragraph{$\text{ABC}^{\text{SVM}}_{\text{PM}}$} The $\text{ABC}^{\text{Grid}}_{\text{PM}}$ is only based on the raw prepaid grid points. But again, a more accurate estimation can be found by interpolating between the parameters in the prepaid grid. Therefore, we learn the relation between the parameters and the summary statistics using LS-SVM: $\boldsymbol{\hat{f}}_{svm}: \boldsymbol{\theta} \mapsto \boldsymbol{s}$. We only learn this relation in the region of interest, that is, only the 100 nearest neighbors according to the $\text{ABC}^{\text{Grid}}_{\text{PM}}$ approach or more specifically, the 100 prepaid points for which the most samples lead to a small enough $\epsilon_{q,i,T_{\mathrm{ABC}}}$.

\begin{figure}[ht]
  \centering
    \includegraphics[width=1\textwidth]{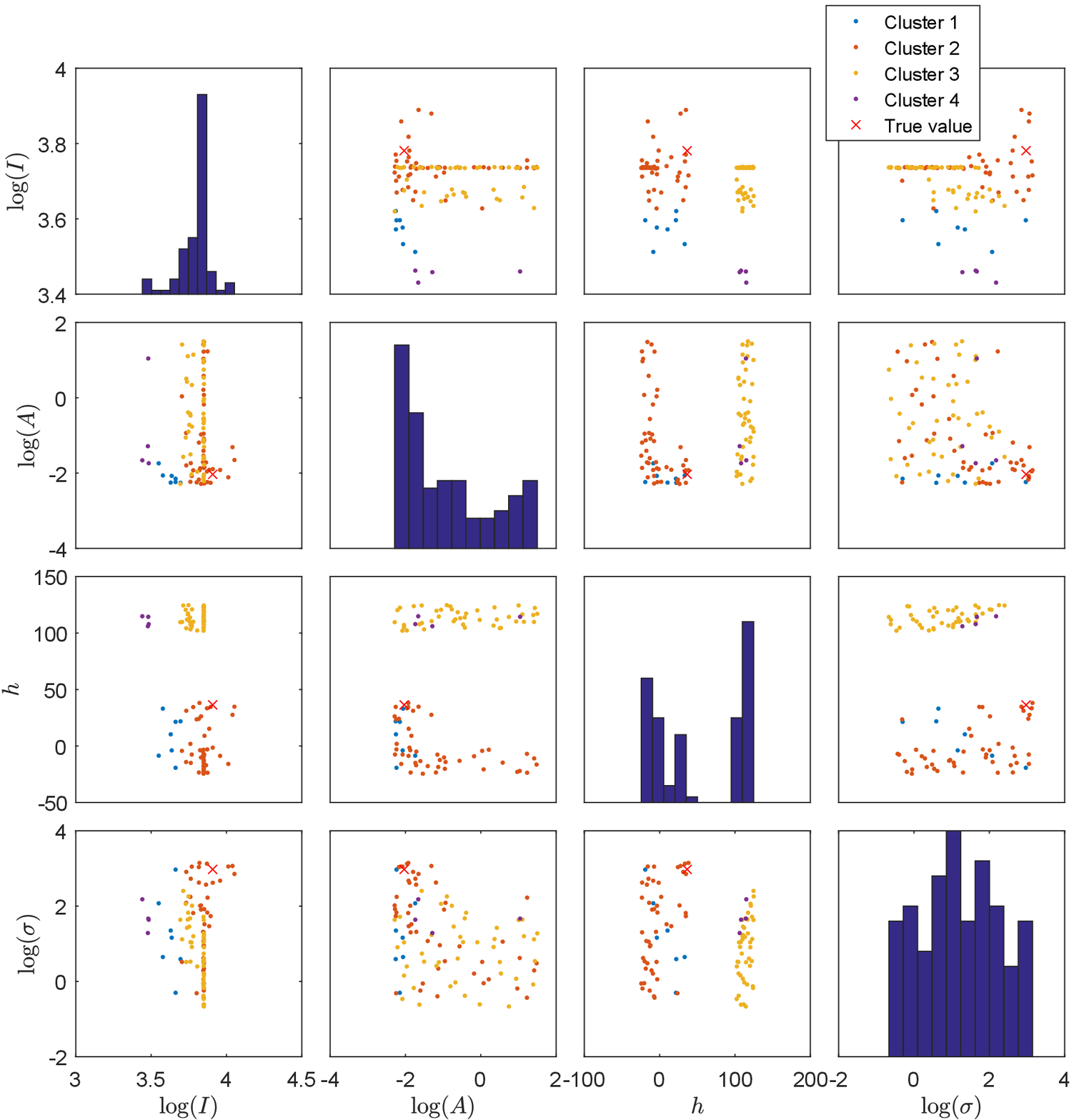}
     \caption{Scatter plot matrix of the clustering that occurs for the 100 nearest neighbors for the summary statistics for $T_{\mathrm{obs}}=1000$ of parameter $\log (I)=3.9081$, $\log(A)=-2.0343$, $h=36.4150$ and $\log(\sigma)=2.9762$. The red cross shows the true value of this parameter.}
     \label{traitCluster}
\end{figure}

Before we use machine learning to infer the relation $\boldsymbol{\hat{f}}_{svm}: \boldsymbol{\theta} \mapsto \boldsymbol{s}$ we cluster these 100 nearest neighbors using hierarchical clustering such that no cluster has more than 50 prepaid points. This is necessary as these 100 nearest neighbors may come from totally different areas in the prepaid grid. This is illustrated in Figure \ref{traitCluster}.

For each cluster, we first make sure that at least 20 points are included (if not, we add points from the prepaid grid which are closest). Then we estimate the $\boldsymbol{\hat{f}}_{svm}: \boldsymbol{\theta} \mapsto \boldsymbol{s}$ using LS-SVM for each cluster $c$ separately, giving rise to $\boldsymbol{\hat{f}}_{svm,c}$. Next, we find the minimum volume ellipse encompassing all the points in each cluster. These ellipses inform us about the areas for which the relation holds. Subsequently we resample parameters in each ellipse to zoom in more and more to the regions of interests. In detail, we do the following in every cluster $c$:
\begin{enumerate}
\item Uniformly sample 1000 points $\boldsymbol{\theta}_{j,c}$ in the minimum volume ellipse for cluster $c$. We create a finer grid for each elipse.
\item Find the summary statistics based on the LS-SVM in cluster $c$: $\hat{\boldsymbol{s}}_{j,c}=\boldsymbol{\hat{f}}_{svm,c}({\boldsymbol{\theta}}_{j,c})$
\item Find for each point $\boldsymbol{\theta}_{j,c}$ the nearest point $\boldsymbol{\theta}_{p}$ from the prepaid points with which this particular cluster was created
\item Translate the 1000 samples from the nearest point $\boldsymbol{\theta}_{p}$ to the newly sampled point $\boldsymbol{\theta}_{j,c}$ and add to each sample the difference in summary statistics: $\boldsymbol{d}=\hat{\boldsymbol{s}}_{j,c}-\boldsymbol{s}_p$. In this step we aproximate a distribution of statistics for $\boldsymbol{\theta}_{j,c}$ around $\hat{\boldsymbol{s}}_{j,c}$.
\item Keep the points $\boldsymbol{\theta}_{j}$ for which $\epsilon_{j,i}$ from Equation \ref{epsTrait} is among the 5000 smallest distances and remove all others.
\item Recalculate the minimum volume ellipse with the new points.
\item Go back to step 1, until the worst $\epsilon_{j,i}$ does not decrease any more.
\end{enumerate}

Broadly speaking, in step 1, we sample parameters $\boldsymbol{\theta}_{j,c}$, in step 2 to 4 we approximate the summary statistics distribution for each $\boldsymbol{\theta}_{j,c}$ using LS-SVM and in step 5 to 7 we trim this set of parameters to only keep the parameters with a high posterior probability.

In the end we combine all the samples, we build the posterior with the parameters from the 1000 best samples over all clusters according to Equation~\ref{epsTrait}. Note that some parameters may show up several times in this posterior sample. To compute the posterior mean, we use a weighted version of these samples. The weights are given by the volume of the ellipse from the cluster where they were created. This is necessary to ensure the correct use of the equal prior for all clusters.

\subsubsection*{Test set}
To generate the test set, we follow the same logic as in \cite{jabot_stochastic_2010}. We use the prior in Equation~\ref{traitPrior} to generate 1000 random parameter sets, except for $h$, where we changed the prior with the following generating distribution:

\begin{equation}
h \sim \mathcal{U}\left(0,100\right),
\end{equation} 
such that 0 and 100 are the true minimum and maximum optimal trait values for communities. By taking the prior for $h$ as in Equation~\ref{traitPrior}, we avoid boundary effects. To exclude  other problems at the borders of the parameter space, we deleted parameters which where within 1\% range of the bounds. We simulated data sets for both $T_{\mathrm{obs}}=1$ and $T_{\mathrm{obs}}=1000$. 

\subsubsection*{Results}
\paragraph{Accuracy}
Let us first look at the results for $T_{\mathrm{obs}}=1$. We have used traditional ABC ($\text{ABC}^{\text{Orig}}$), prepaid Bayes approach based on the synthetic likelihood ($\text{SL}^{\text{Grid}}_{\text{PM}}$) and prepaid ABC based on separately generated samples at the grid points ($\text{ABC}^{\text{Grid}}_{\text{PM}}$ and $\text{ABC}^{\text{SVM}}_{\text{PM}}$). We have used $10^5$ and $5\cdot 10^5$ prepaid grid points. The RMSE and MAE can be found in Tables~\ref{tabap2rmse} and~\ref{tabap2mae}. All methods result in accuracies that are equally large. For 3 out of 4 parameters (except for $h$), the prepaid method outperforms $\text{ABC}^{\text{Orig}}$ with respect to RMSE. For MAE, the prepaid method uniformly outperforms the Easy ABC package ($\text{ABC}^{\text{Orig}}$). Overal, the difference between $\Omega=10^5$ and $\Omega=5\cdot 10^5$ prepaid grid point is very small for the prepaid methods.

We have refrained from interpolating with the LS-SVM because the $99.9\%$ coverage includes on average more than 1000 points. This is perfectly logical because $T_{\mathrm{obs}}=1$ does not provide a lot of information, and, as a consequence, there is a lot of uncertainty (which translates itself into a large number of parameter points that have a reasonable large synthetic likelihood value). As a result, creating a posterior based on only 100 nearest neighbors (even after interpolation) would not suffice because too many parameter points with high posterior density would be missed.  

For $T_{\mathrm{obs}}=1000$ (see again Tables~\ref{tabap2rmse} and~\ref{tabap2mae}), the accuracy increases, as is expected (this can be seen both in the RMSE as in the MAE). In this case, both increasing the number of grid points $\Omega$ and using LS-SVM interpolation increases accuracy. No results are given for $\text{ABC}^{\text{Orig}}$, because it is impossible to fit the model with  this sample size in acceptable time.

\paragraph{Speed}
For $T_{\mathrm{obs}}=1$, the estimation time of $\text{ABC}^{\text{Orig}}$ is 3865 s. In contrast, the estimation time of $\text{ABC}^{\text{Grid}}_{\text{PM}}$ is 0.167 s. This means that the prepaid ABC method is approximately 23000 times faster than traditional ABC.

\paragraph{Coverage}
For both the $\text{ABC}^{\text{Orig}}$ as well as the prepaid versions we end up with a posterior sample. We computed the coverage by calculating the 0.025 and 0.975 quantiles of the posterior samples. Next, we checked whether the true parameter was in this interval or not. Note that when we use clustering during $\text{ABC}^{\text{SVM}}_{\text{PM}}$, we weigh each point proportional to the volume of its originating cluster. For the $\text{SL}^{\text{Grid}}_{\text{PM}}$ approach we use the whole prepaid set as posterior and us weights according to Equation~\ref{eqPSL}.

For $T_{\mathrm{obs}}=1$ and $T_{\mathrm{obs}}=1000$, coverage results can be found in Table~\ref{tabap2cov}. For  $T_{\mathrm{obs}}=1$, $\text{ABC}^{\text{Orig}}$ leads to better coverages than  $\text{SL}^{\text{Grid}}_{\text{PM}}$. Also the $\text{ABC}^{\text{Grid}}_{\text{PM}}$ method gives good coverages (around the nominal level of 0.95) for $T_{\mathrm{obs}}=1$, but these coverages deteriorate for $T_{\mathrm{obs}}=1000$ if no interpolation is used (coverage is a bit better for $5\cdot 10^5$ grid points). When the LS-SVM interpolation is applied (i.e., $\text{ABC}^{\text{SVM}}_{\text{PM}}$), coverages become very good again, certainly for the largest number of grid points.

\begin{table}[H]
\caption{The MAE of the estimations of the test set of the trait model.}
\begin{center}
\begin{tabular}{c l l| r r r r}
 $T_{\mathrm{obs}}$ & version & $\Omega$ & $\log (I)$ & $\log(A) $ & $h$  & $\log(\sigma)$\\ \hline
1 & $\text{ABC}^{\text{Orig}}$ & /  & 0.11 & 0.45 & 1.4 & 0.45\\
1 & $\text{SL}^{\text{Grid}}_{\text{PM}}$ & 100000   & 0.1 & 0.39 & 0.96 & 0.38\\
1 & $\text{ABC}^{\text{Grid}}_{\text{PM}}$ & 100000   & 0.1 & 0.4 & 1 & 0.4 \\
1   & $\text{ABC}^{\text{Grid}}_{\text{PM}}$ & 500000  &  0.1 & 0.38 & 1 & 0.39 \\ \hline
1000 & $\text{ABC}^{\text{Grid}}_{\text{PM}}$ & 100000  & 0.03 & 0.14 & 0.39 & 0.32  \\
1000 & $\text{ABC}^{\text{Grid}}_{\text{PM}}$ & 500000   & 0.02 & 0.09 & 0.27 & 0.22 \\
1000 & $\text{ABC}^{\text{SVM}}_{\text{PM}}$ & 100000   & 0.02 & 0.07 & 0.18 & 0.14  \\
1000 & $\text{ABC}^{\text{SVM}}_{\text{PM}}$ & 500000   & 0.01 & 0.07 & 0.17 & 0.15  \\
\end{tabular}
\end{center}
\label{tabap2mae}
\end{table}

\begin{table}[H]
\caption{The effective 95$\%$ coverage of the estimations of the test set of the trait model.}
\begin{center}
\begin{tabular}{c l l| r r r r}
 $T_{\mathrm{obs}}$ & version & $\Omega$  & $\log (I)$ & $\log(A) $ & $h$  & $\log(\sigma)$\\ \hline
1 & $\text{ABC}^{\text{Orig}}$  & /  & 0.97 & 0.97 & 0.99 & 0.96\\
1 & $\text{ABC}^{\text{Orig}}$ & 100000  & 0.84 & 0.87 & 0.86 & 0.86\\
1 & $\text{ABC}^{\text{Grid}}_{\text{PM}}$ & 100000  & 0.94 & 0.95 & 0.95 & 0.94 \\
  1 & $\text{ABC}^{\text{Grid}}_{\text{PM}}$ & 500000  & 0.94 & 0.95 & 0.94 & 0.94 \\ \hline
1000 & $\text{ABC}^{\text{Grid}}_{\text{PM}}$ & 100000   & 0.27 & 0.3 & 0.29 & 0.27 \\
1000 & $\text{ABC}^{\text{Grid}}_{\text{PM}}$ & 500000 & 0.47 & 0.5 & 0.48 & 0.48  \\
1000 & $\text{ABC}^{\text{SVM}}_{\text{PM}}$ & 100000  & 0.93 & 0.94 & 0.96 & 0.93  \\
1000 & $\text{ABC}^{\text{SVM}}_{\text{PM}}$ & 500000 & 0.96 & 0.95 & 0.96 & 0.95  \\
\end{tabular}
\end{center}
\label{tabap2cov}
\end{table}

%\begin{figure}[ht]
%  \centering
%    \includegraphics[width=1\textwidth]{easyABC.eps}
%     \caption{The improvement of the Easy ABC algorithm of all parameters over the decreasing tolerance %parameter $\rho$. The blue lines are $\log (I)$, the red lines $\log(A)$, the yellow lines $h$ and the purple %$\log(\sigma)$.}
%     \label{easyABC}
%\end{figure}
% figuur wordt niet echt gebruikt

\subsection{Application 3: The Leaky Competing Accumulator} \label{subs:lca}
Elementary decision making has been studied intensively in humans and animals \cite{huk_role_2017}. A common example of an experimental paradigm is the random-motion dot task: the participant has to decide whether a collection of dots (of which only a fraction moves coherently; the others move randomly) is moving to the left or to the right. The stimuli typically have varying levels of difficulty, determined by the fraction of dots moving coherently.

Assuming there are two response options (e.g., left and right), the Leaky Competing Accumulator consists of two evidence accumulators, $x_{1}(t)$ and $x_{2}(t)$ (where $t$ denotes the time), each associated with one response option. The evolution of evidence across time for a single trial is then described by the following system of two stochastic differential equations:

\begin{equation} \label{eq:lcamodel}
\begin{split}
dx_{1}   &= (v+\Delta v_{i}-\gamma x_{1}-\kappa x_{2}) \cdot dt+c \cdot dW_{1}\\
dx_{2}   &= (v-\Delta v_{i}-\gamma x_{2}-\kappa x_{1}) \cdot dt+c \cdot dW_{2},
\end{split}
\end{equation}
where $dW_{1}$ and $dW_{2}$ are uncorrelated white noise processes. To avoid negative values, the evidence is set to 0 whenever it becomes negative: $x_{1}  =  \max(x_{1},0)$ and $x_{2}  = \max(x_{2},0)$. The initial values (at $t=0$) are $(x_{1},x_{2})=(0,0)$.

The evidence accumulation process continues until one of the accumulators crosses a boundary $a$ (with $a>0$). The coordinate that crosses its decision boundary first, determines the choice that is made and the time of crossing is seen as the decision time. The observed choice response time is seen as the sum of the decision time and a non-decision time $T_{er}$, to account for the time needed to encode the stimulus and emit the response. 

Equation~\ref{eq:lcamodel} describes the evolution of information accumulation for a two-option choice RT task, given the presentation of a single stimulus. For all stimuli, the total evidence is equal to $v$, but the differential evidence for option 1 compared to 2 is $2\Delta v_i$, which is stimulus dependent and reflects the stimulus difficulty. In this example, we assume the stimuli can be categorized into four levels of difficulty, hence $i=1,\dots,4$. 

The model gives rise to two separate choice response time probability densities, $p_{1i}(t)$ and $p_{2i}(t)$, each representing the response time conditional on the choice that was made. Integrating the densities over time will result in the probability of choosing the response options: $\int_0^{\infty} p_{1i}(t) dt = \Pr(\mbox{option 1 for stimulus $i$})$ and $\int_0^{\infty} p_{2i}(t) dt = \Pr(\mbox{option 2 for stimulus $i$})$. Obviously, when taken together, $p_{1i}$ and $p_{2i}$ sum to one. 

All parameters in the parameter vector $\boldsymbol{\theta}=(v,\Delta v_1,\dots,\Delta v_4,\kappa,\gamma,a,T_{er})$ can take values from $0$ to $\infty$. This parametrization is known to have one redundant parameter \cite{miletic_parameter_2017}, so we choose to fix $c=0.1$.

\subsubsection*{The re-parametrization}
The prepaid method will not be applied to the model as presented in Equation~\ref{eq:lcamodel}, but rather on a re-parametrized formulation:

\begin{equation} \label{eq:lcarepar}
\begin{split}
dx_{1it} & =  D\cdot\left(v'(1+C_i)-\gamma' \cdot x_{1it}-\kappa' \cdot x_{2it}\right)\cdot dt+\sqrt{D} \cdot dW_{1it}\\
dx_{2it} & =  D\cdot\left(v'(1-C_i)-\gamma' \cdot x_{2it}-\kappa' \cdot x_{1it}\right)\cdot dt+\sqrt{D} \cdot dW_{2it},
\end{split}
\end{equation}
again with the additional restriction that $x_{1it}  =  \max(x_{1it},0)$ and $x_{2it}  = \max(x_{2it},0)$. The new parameters are defined as follows in terms of the original ones:

\begin{eqnarray*}
D & = & c^{2}\\
v' &=& \frac{v}{D}\\
C_i &=& \frac{\Delta v_i}{v}\\
\gamma' & = & \frac{\gamma}{D}\\
\kappa' & = & \frac{\kappa}{D}.
\end{eqnarray*}
This new parametrization has the advantage that $D$ can be interpreted as an inverse time scalar because doubling $D$ makes all choice response times twice as fast. This property will allow us to reduce the dimensionality of the prepaid grid (see below). The parameter $v'>0$ denotes general stimulus strength scaled according to $D$, while parameter $C_i$ (for coherence) denotes the amount of relative evidence encoded in the stimulus $i$: $-1<C_i<1$. It is commonly assumed for these evidence accumulator models that different stimuli should lead to different coherences $C_i$, but without affecting the other parameters. The nondecision time $T_{er}$ is not transformed.

%In terms of the original parametrization, this means that in addition to the parameters that are not directly related to the input, also the sum of the inputs $v_{1},v_{2}$ should remain constant across stimuli. 

% with a noise injection to de-bias the estimates->niet in hoofdtekst, wordt hier wel gezegd=ok
%ergens zeggen hoeveel prepaid punten ge gebruikt voor LCA

\subsubsection*{Creation of the prepaid grid}
For the delineation of the parameter space, we will follow the specifications of \cite{miletic_parameter_2017}. Because this parameter space is rather restrictive (a consequence of the recommendation of \cite{miletic_parameter_2017} to improve parameter recovery), we will extend it through the use of a time scale parameter. This extension will be further discussed when introducing the test set.

First, we create a prepaid grid on a four-dimensional space in the original parametrization by drawing from the following distribution:

\begin{equation}
\begin{split}
a &\sim \mathcal{U} (0.05,0.25) \\ 
v &\sim \mathcal{U} (0.8,1.5) \\
\gamma &\sim \mathcal{U} (1,8) \\
\kappa &\sim \mathcal{U} (1,8).
\end{split}
\label{lca:prior1}
\end{equation}

We select 10000 grid points from this distribution using Halton sequences \cite{matlab_version_2016, kocis_computational_1997}. When working in the reparametrized version, as defined in Equation~\ref{eq:lcarepar}, this space can be transformed to a four dimensional space of $v'$, $\gamma'$, $\kappa'$ and $D$.

\begin{figure}[ht]
  \centering
  \includegraphics[width=12cm]{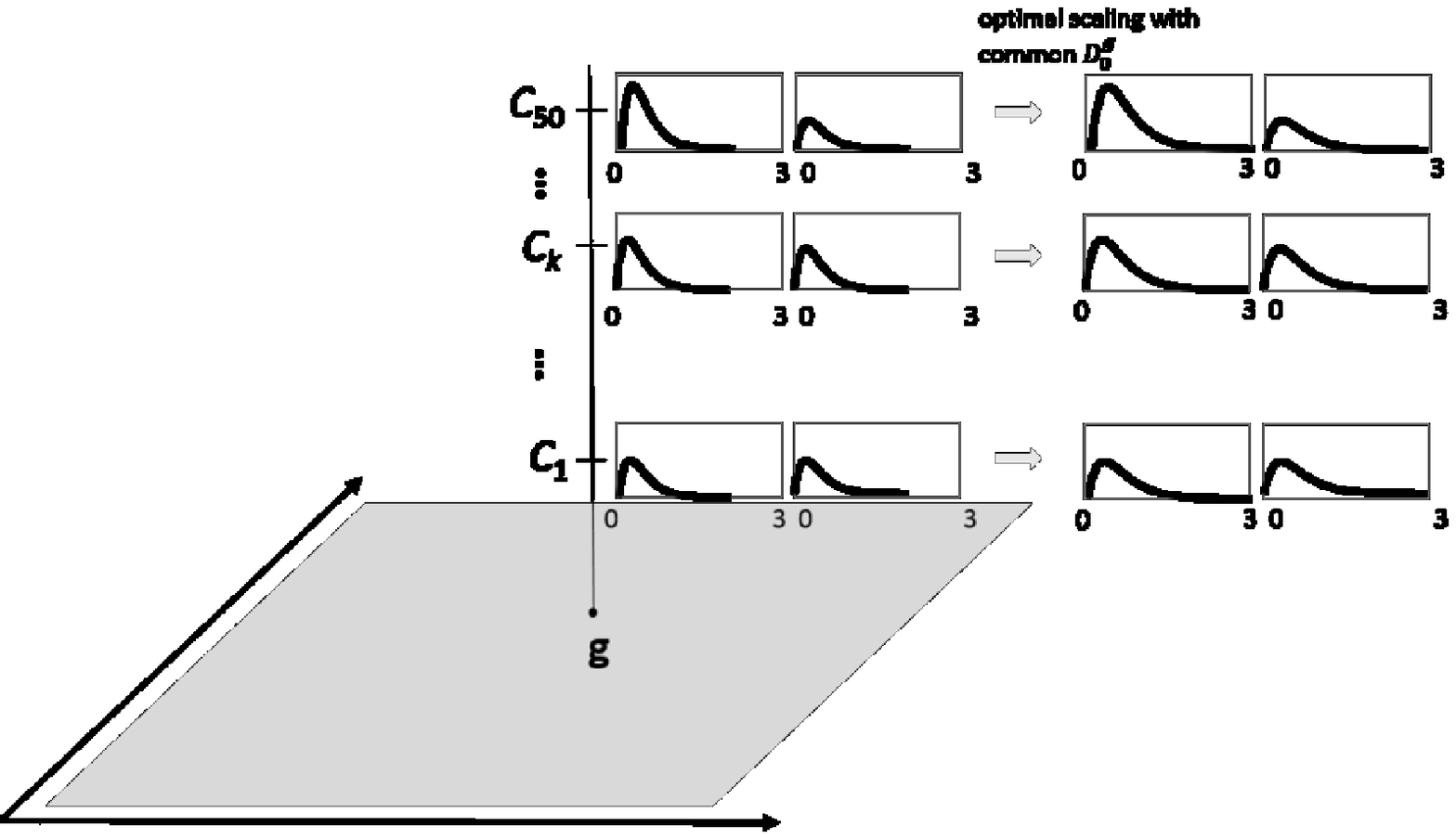}
     \caption{Illustration of how different coherences are incorporated. The gray plane is a simplified representation of the three dimensional $(v', \gamma', \kappa')$-space. For each point $\boldsymbol{g}$, 50 coherences are chosen. Corresponding to each coherence, there is a pair of RT distributions (which each integrate to the probability of selecting the corresponding option).}
     \label{fig:lcacoh}
\end{figure}

\begin{figure}[ht]
  \centering
  \includegraphics[width=12cm]{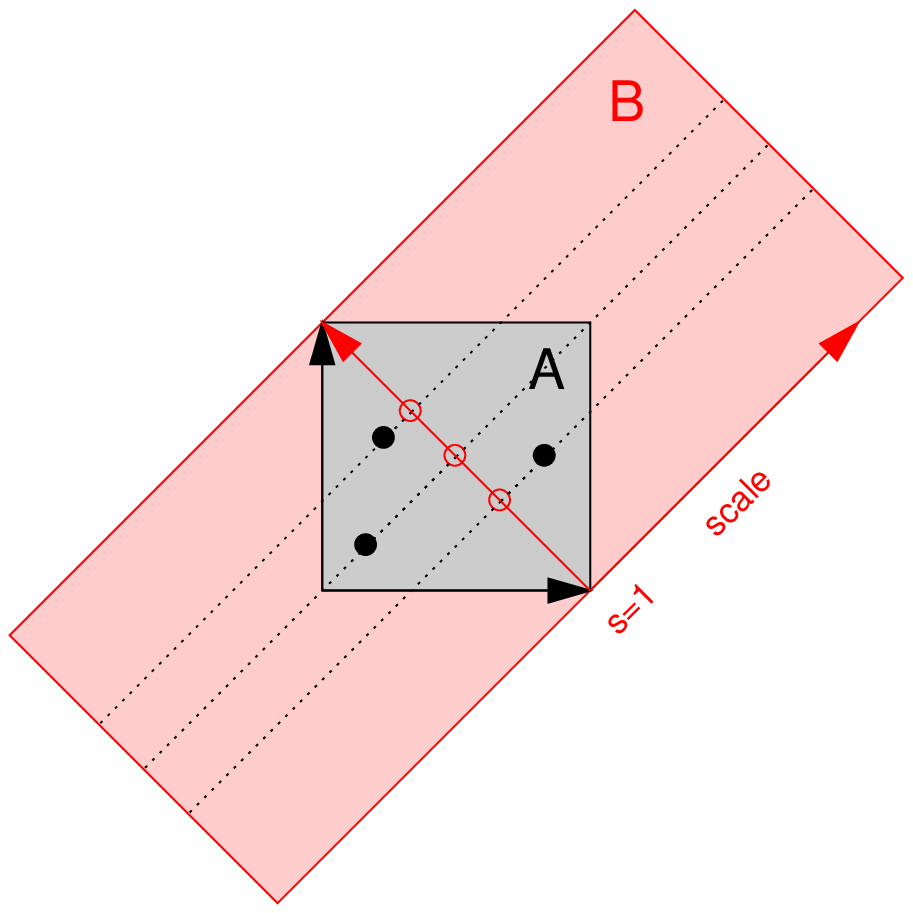}
     \caption{Illustration of the transformation of the original parameter space (called $A$) to a new one (called $B$) in which $D$ is one of the parameters. The projections of the three parameter points on the red axis governing the width of the $B$ area are denoted with open circle and these are the parameter points $\boldsymbol{g}$. For each of these open circle points, the RT distribution scales are set to 1 (i.e., $s=1$) by choosing an appropriate diffusion coefficient (denoted as $D_0^{\boldsymbol{g}}$) and any parameter point in $B$ can be reached by selecting an appropriate $\boldsymbol{g}$ and then adjusting the scale up- or downwards (this is indicated by the dotted lines in the length direction of the new parameter space $B$.}
     \label{fig:parspace}
\end{figure}

However, because $D$ acts an inverse time scalar on the response time distributions, we may also consider the three dimensional space formed by $v'$, $\gamma'$, and $\kappa'$ and for each grid point, choose the parameter $D$ in such a way that the RT distributions for options 1 and 2 are scaled to fit nicely between 0 and 3 seconds (with a resolution of 1ms and 3000 time points so that about $0.0001$ of the tail mass is allowed to be clipped at 3 seconds when $C=0$). Effectively, this brings all RT distributions to the same scale (denoted as $s=1$). This process of scaling is illustrated in Figure~\ref{fig:lcacoh}. It reduces both the number of simulations and the storage load (without it we would have to simulate and store a separate set of distributions for each value of $D$). Note that the scaling is done jointly for all RT distributions associated with a particular $\boldsymbol{g}$. The resulting diffusion constant corresponding to the rescaled distribution is denoted as $D_0^{\boldsymbol{g}}$. In addition, the construction effectively removes one parameter from the prepaid grid, which is illustrated in Figure~\ref{fig:parspace}.

To include the coherence parameter, we extend each grid point with a set of predefined coherences. For each point $\boldsymbol{g}=(v', \gamma', \kappa')$ in the grid, we take 50 equally spaced coherences $C^{\boldsymbol{g}}_k$ (with $k=1,\dots,50$) from $0$ to the maximum coherence that still has some non-zero chance of choice option 2 to be selected (we take $0.001$). Finally, we simulate for each combination of $\boldsymbol{g}=(v', \gamma', \kappa')$ and $C^{\boldsymbol{g}}_k$ a large number of choice response time data (choices and response times). This is illustrated in Figure~\ref{fig:lcacoh}.

In a last step, grid points are eliminated from the prepaid grid, if the simulations result in too many simultaneous arrivals (i.e., trajectories that end at or very close to the intersection point of the two absorbing boundaries at the upper right corner, located at $(a,a)$). More specifically, we drop grid points with more than 0.1 percent simultaneous arrivals. Creating the prepaid database took less then a day on a NVIDIA GeForce GTX 780 GPU.

\subsubsection*{Prepaid estimation}

%\subsubsection*{Stimulus selection and integrating out shift and time scale}

To explain how the prepaid estimation of the LCA works, let us start with a prototypical experimental design. Assume a choice RT experiment with four stimulus difficulty levels (e.g., four coherences in the random dot motion task). Each difficulty level is administered $N$ times to a single participant. A particular trial in this experiment results in $(c_{ij},t_{ij})$, where $i$ is the stimulus difficulty level ($i=1,\dots,4$) and $j$ is the sequence number within its difficulty level ($j=1,\dots,N$). The data resulting from this experiment are responses $c_{ij}$ (referring
to choice $1$ or choice $2$) and response times $t_{ij}$. Each pair $(c_{ij},t_{ij})$ is considered to originate from an unknown parameter set $(v',\gamma',\kappa',D,T_{er})$ and coherences $C_{i}$ ($i=1,\dots,4$).

Our first aim is to is to establish a local net of prepaid points that lead to data that are close to the observed dataset. If necessary, we can further zoom in with the help of support vector machines. Conditional on each prepaid parameter set $\boldsymbol{g}$ in the basic grid, a number of the remaining parameters can be integrated out beforehand. First, conditional on grid point
$\boldsymbol{g}$, we have for 50 predetermined coherences $C^{\boldsymbol{g}}_k$ simulated accuracies and response time distributions (see Figure~\ref{fig:lcacoh}). The coherences of the observed data can be estimated solely using the observed accuracies using simple linear interpolation. The estimated coherence for stimulus (or condition) $i$ is denoted as $\hat{C}_{i}$. Corresponding to each of the 50 coherences $C^{\boldsymbol{g}}_k$ for grid point $\boldsymbol{g}$, there is a pair of corresponding simulated RT densities $p_{ic}^{\boldsymbol{g}}(t)$ (with $c=1,2$). As before, $p_{ic}^{\boldsymbol{g}}(t)$ is scaled to the $[0,3]$ seconds window, and we can use a combination of translating (estimating $\hat{T}_{er}$), scaling (estimating $\hat{D}$) and interpolating. Specifically, we first calculate $\hat{s}$ as the optimal time scalar to match data with the model on grid point $\boldsymbol{g}$:

\begin{equation*}
\hat{s} = \sqrt{\frac{\frac{1}{(4N-1)}\sum_{ij}(t_{ij}-\hat{\mu}_{t})^{2}}{\frac{1}{4}\sum_{ic}\int p_{ic}^{\boldsymbol{g}}(t)(t-\mu_{t}^{\boldsymbol{g}})^{2}dt}},
\end{equation*}
in which

\begin{eqnarray*}
\hat{\mu}_{t} & = & \frac{1}{4N}\sum_{ij}t_{ij}\\
\mu_{t}^{\boldsymbol{g}} & = & \frac{1}{4}\sum_{ic}\int p_{ic}^{\boldsymbol{g}}(t)tdt.
\end{eqnarray*}
This formula capitalizes on the fact that the variance of a distribution does not change when it is simply shifted to the right by a constant. Hence, the ratio of the model's decision time variance (without $T_er$) and the observed total response time variance (presumably shifted with some $T_er$) is still an estimator of the squared scale factor between them.
Using this information, we can estimate the optimal $\hat{D}$ and $\hat{T}_{er}$ for grid point $\boldsymbol{g}$ as follows: 

\begin{eqnarray*}
\hat{D} & = & \frac{D_{0}^{\boldsymbol{g}}}{\hat{s}}\\
\hat{T}_{er} & = & \hat{\mu}_{t}-\hat{s}\mu_{t}^{\boldsymbol{g}},
\end{eqnarray*}
with $D_{0}^{g}$ being the optimal scaling diffusion constant used for optimal storage in the database. This gives us a final effective
parameter vector of $\left(v',\gamma',\kappa',\hat{C}_{1},\hat{C}_{2},\hat{C}_{3},\hat{C}_{4},\hat{D},\widehat{T}_{er}\right)$.
Note that the last 6 elements of this vector are estimates conditional on the grid point $\boldsymbol{g}=(v',\gamma',\kappa')$. 

Next, we have to determine the single optimal parameter set (and thus also the optimal $v'$, $\gamma'$, and $\kappa'$). For this we need an objective function that compares the model based PDFs with those of the data. For this purpose, we use a (symmetrized) chi-square
distance based on a set of bin statistics. For each stimulus' observed set of choice RTs, $\boldsymbol{t}_{i}=(\boldsymbol{t}_{i1},\boldsymbol{t}_{i2})$ (with $\boldsymbol{t}_{i1}$ the RTs for option 1 and $\boldsymbol{t}_{i2}$ for option 2), we calculate 20 data quantiles $q_{u}$  (with $u=1,\dots,20$) at probability masses $m_i=0.05\cdot i$. The set of quantiles is appended with one extra quantile $q_0$ at $m_0=0.01$ to have a more detailed representation of the leading edge of the distribution. Based on binning edges $\left(0,q_0,q_1,\dots,q_{20},+\infty\right)$, we create $4\times2\times22$ bin frequencies $\hat{b}_{icw}$ with $w=1,\ldots,22$.
The corresponding probability masses $m_{icw}^{g}$ can be easily
extracted from the prepaid PDFs $p_{ic}^{g}(t)$ as well. Observed and theoretical quantities can then be combined in the a symmetrized chi-square distance:

\begin{eqnarray} \label{eq:objfun_lca}
d(g,\{c_{ij},t_{ij}\}) & = & \sum_{icw}\frac{(\hat{b}_{icw}-m_{icw}^{g}){}^{2}}{\hat{b}_{icw}+m_{icw}^{g}}
\end{eqnarray}
This defines a distance between all grid points $g$ in the database
and any data set. 

In the following paragraphs we will present three ways of using this distance to calculate LCA estimates, each a bit more complicated than the previous one (but also more accurate): $\text{CHISQ}^{\text{Grid}}_{\text{NN}}$, $\text{CHISQ}^{\text{Grid}}_{\text{BS}}$, $\text{CHISQ}^{\text{SVM}}_{\text{BS}}$.

\paragraph{$\text{CHISQ}^{\text{Grid}}_{\text{NN}}$}

The grid point closest to the data set (as measured by the symmetrized chi-square distance function) can be used as a first nearest neighbor estimate.

\paragraph{$\text{CHISQ}^{\text{Grid}}_{\text{BS}}$}

Not all parameters are treated equally in the estimation procedure. The parameters $C_i$, $D$ and $T_{er}$ are estimated conditionally on all grid points $\boldsymbol{g}$ and then the other parameters are estimated conditionally on $ \hat{C}_i$, $\hat{D}$ and $\hat{T}_{er}$. Moreover, these parameters are chosen in such a way that a specific aspect of the data (e.g., proportion of choices for option 1) is fitted perfectly (i.e., the coherence is chosen to result in probabilities perfectly equal to the proportions observed in the data). This would be no problem for an infinite amount of data. However, for finite data, the major disadvantage of this way of working is that any errors induced in the precursor step are propagated through the estimation process for $v'$, $\gamma'$ and $\kappa'$. This is because for finite data, the observed accuracies will typically not exactly coincide with
the accuracies provided by the best model estimates. As the estimates $\hat{C}_{i}$ are (on each grid point) exactly fit
to the observed accuracy and consequently, the effective grid points
will all have this exact accuracy. We tackle this estimator bias by non-parametrically bootstrapping the data and repeating the nearest neighbor estimate for every bootstrapped dataset. Taking the mean of this set of estimates (a method known as bagging; \cite{hastie_elements_2009}), gives us a more accurate estimate. Additionally, we now have a standard error of the estimate (and confidence interval).

\paragraph{$\text{CHISQ}^{\text{SVM}}_{\text{BS}}$}

If we apply the bootstrap procedure, it may turn out that the selected grid points as nearest neighbor are not very diverse (this may happen with large sample sizes). In such a situation, it can be worthwhile to use an interpolator. So we may learn a support vector machine based on the bin statistics of the few unique bootstraps grid points available, together with the best overall unique grid points. We propose to use a training set of 100 grid points in total. The SVM can then be used as an approximation for the bin statistics in the space between
the grid points and hence for the objective function. We subsequently minimize the approximative SVM based objective function for every bootstrap, using differential evolution (as has been outlined above for the other applications).

Obviously, the quality of the SVM based estimate is limited by the quality of the SVMs that are trained to learn the relation between parameters and statistics. In addition, the same SVMs are used for all bootstrap samples, which may introduce an unwanted distortion in the uncertainty assessment. To account for the systemic bias that might have been introduced by the SVMs, we will add some random noise to each bootstrap estimate. The amount of random deviation that is added equals the size of the prediction error of the SVM. In this way, low quality SVMs are prohibited of biasing all bootstraps in the same way. The uncertainty of the SVMs is now incorporated in the final bootstrapped results.

\subsubsection*{Test set}

%The test set is uniformly sampled from the prepaid domain. The collapsed parameter D is allowed to vary from 0.5 to 2 times the scale it would have been saved in the 3000 1 millisecond bins prepaid dataset, covering most realistic distributions (REF).

The test set is created by uniformly sampling parameters according to Equation~\ref{lca:prior1}. Input differences $\frac{v_1-v_2}{2}$ are chosen to produce typical accuracies of 0.6, 0.7, 0.8, and 0.9. As is done in \cite{miletic_parameter_2017}, excessively long PDFs (with a maximum RT larger than 5000ms) and excessively short PDFs (with a range below 400ms) are removed from the test set. Apart from the fact that these PDFs are deemed unrealistic \cite{miletic_parameter_2017} for typical choice RT data, this part of the parameter space suffers from inherent poor parameter identifiability, with very large confidence intervals and less meaningful estimates as a consequence. Because the new parametrization analytically integrates out scale (i.e., $D$) (and also shift $T_{er}$), and is positively unbounded in these dimensions, we can expand the test set to cover a broader range of distributions than the ones covered in \cite{miletic_parameter_2017}. To broaden the range of the test, the distributions are scaled with a random factor ranging from 0.2 to 5. We will use this broadened test set to determine the method's accuracy and coverage.

\subsubsection*{Results}

\begin{figure}[ht]
  \centering
    \includegraphics[width=0.5\textwidth]{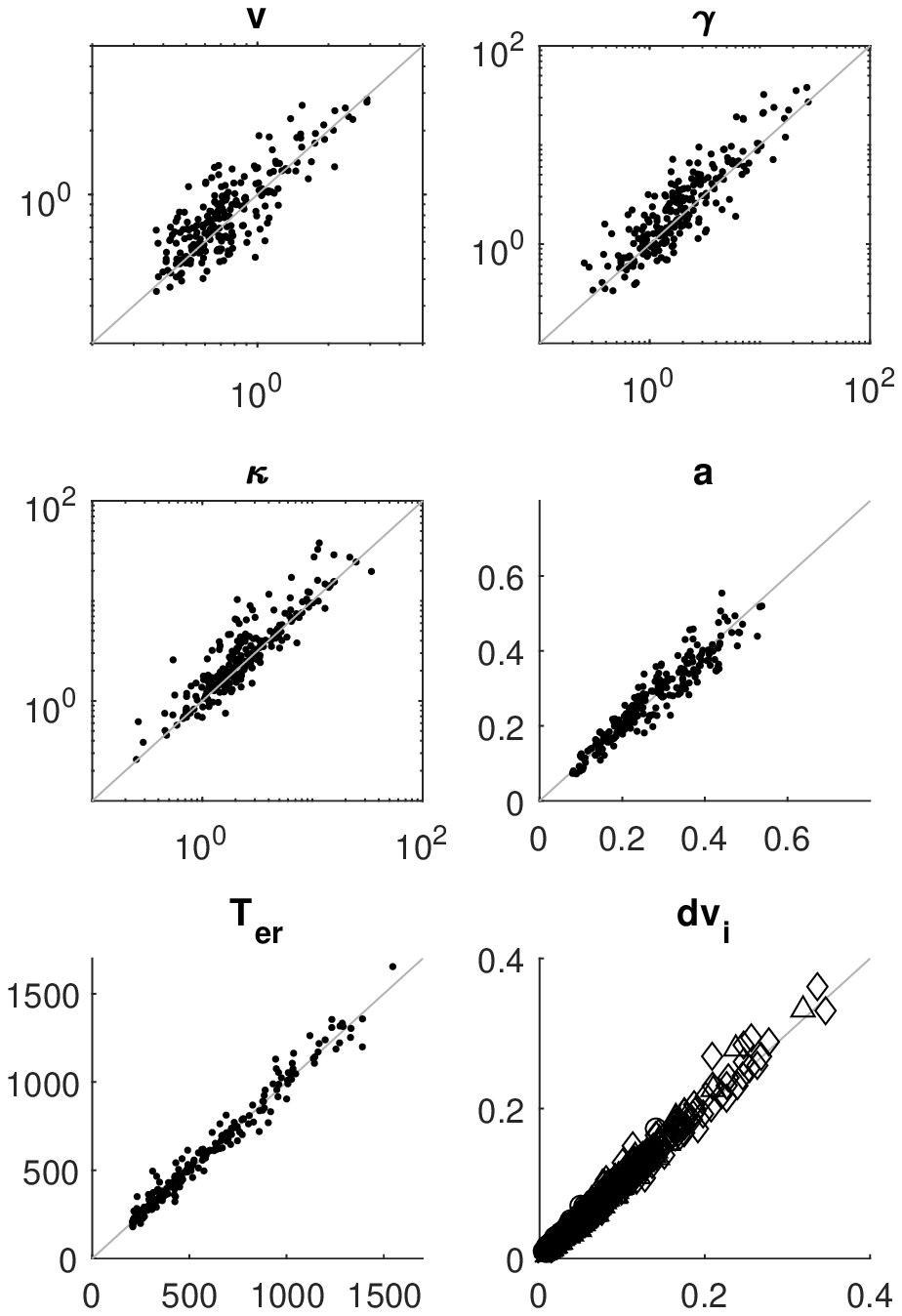}
     \caption{Recovery for the original parameters of the LCA model with $T_{obs}=1000$ observation per stimulus. See Figure~\ref{N300} for detailed information.}
     \label{N1000}
\end{figure}

\begin{figure}[ht]
  \centering
    \includegraphics[width=0.5\textwidth]{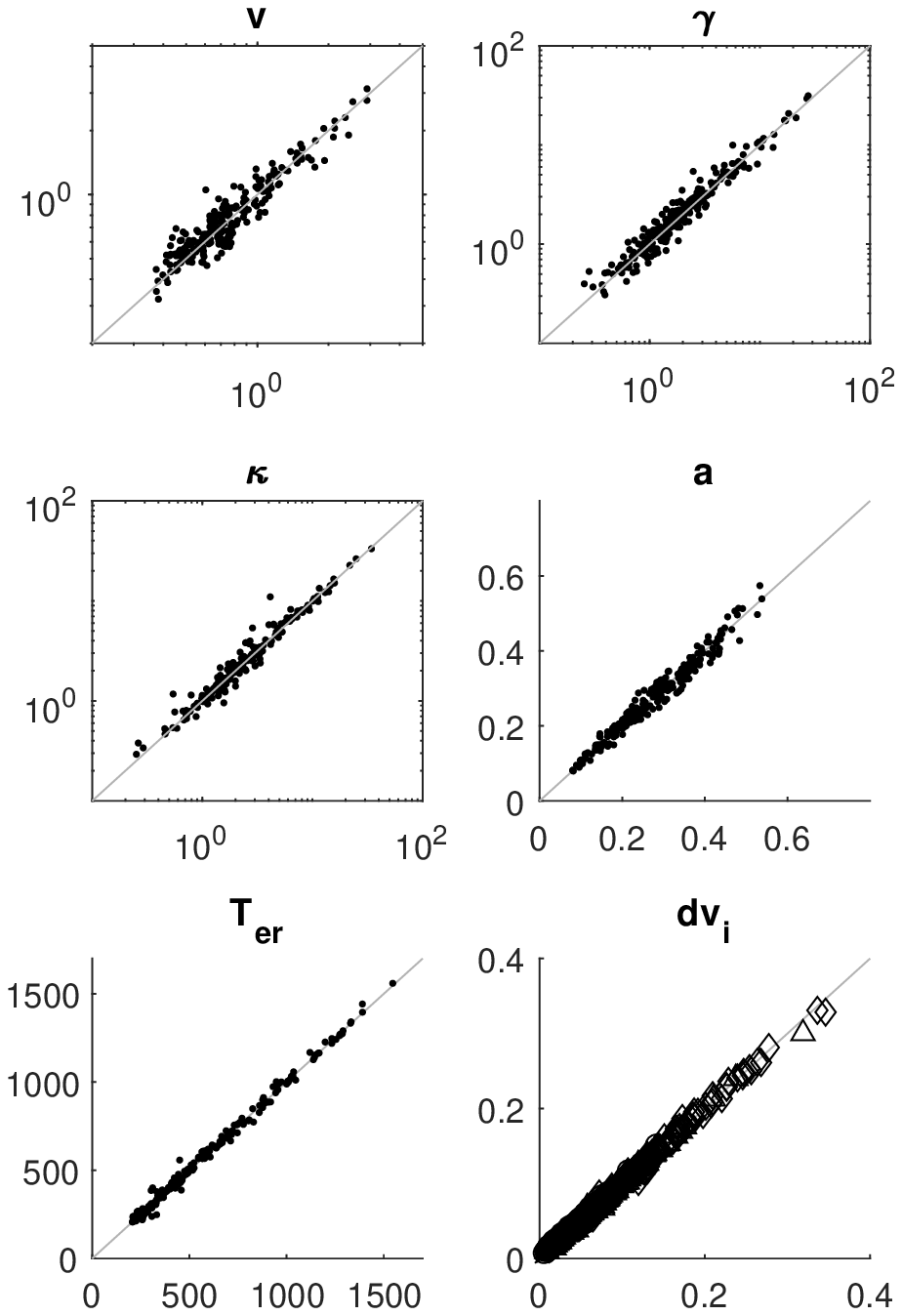}
     \caption{Recovery for the original parameters of the LCA model with $T_{obs}=10000$ observation per stimulus. See Figure~\ref{N300} for detailed information.}
     \label{N10000}
\end{figure}

\begin{figure}[ht]
  \centering
    \includegraphics[width=0.5\textwidth]{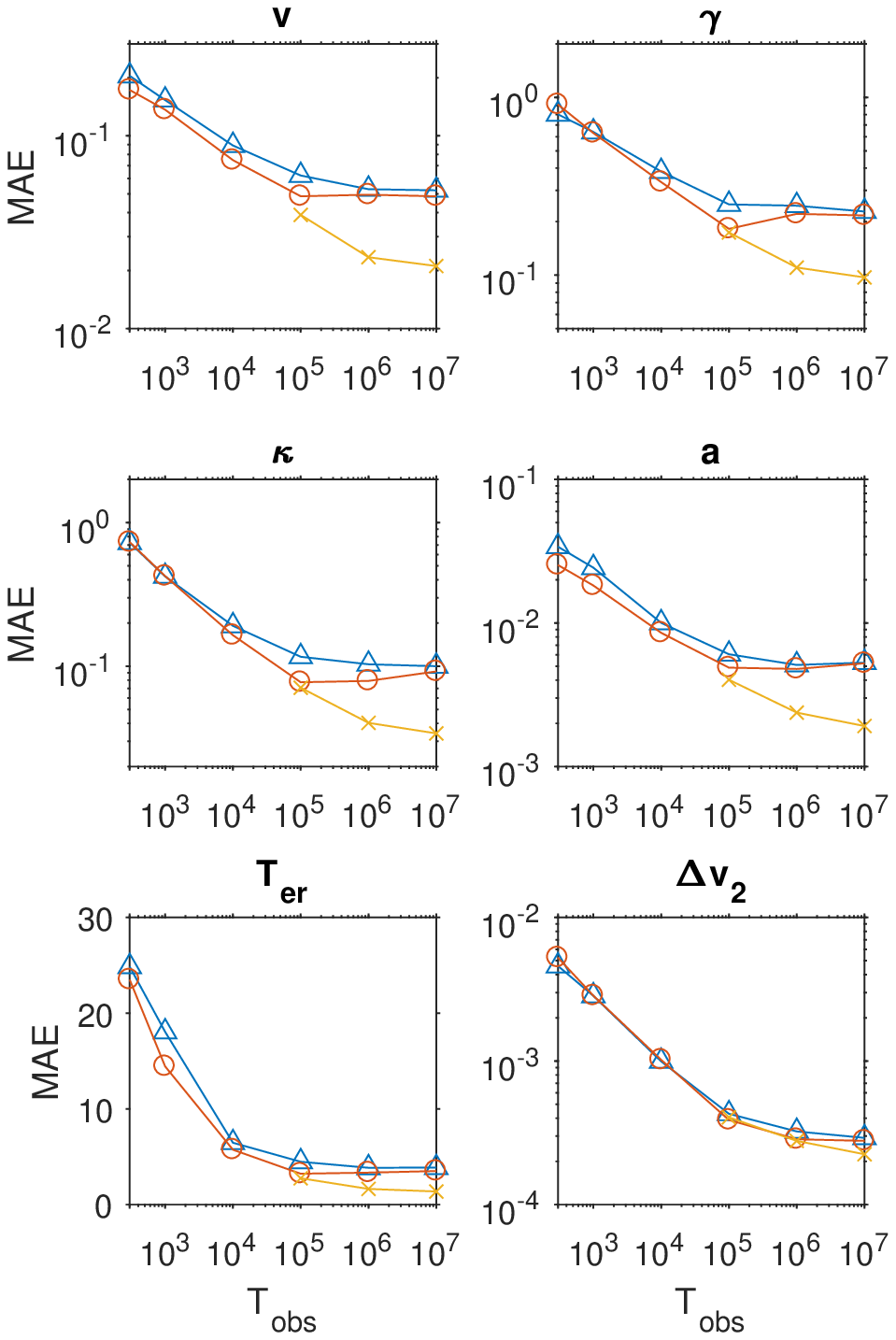}
     \caption{The MAE of the estimates of the parameters of the LCA as a function of sample size (abscissa) and for different methods. More details can be found in the caption of Figure~\ref{mae}.}
     \label{maeAll}
\end{figure}

\begin{figure}[ht]
  \centering
    \includegraphics[width=0.5\textwidth]{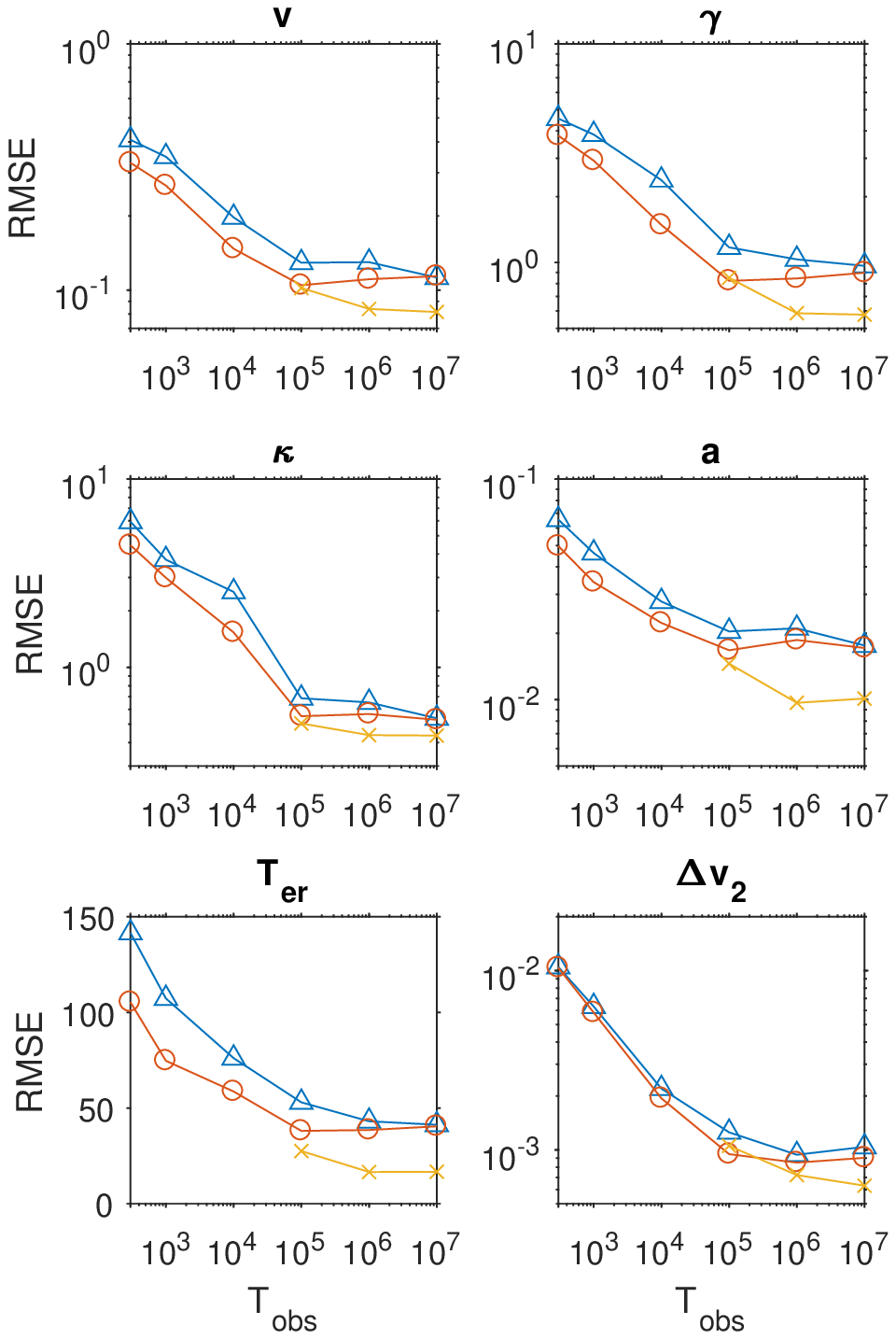}
     \caption{The RMSE of the estimates of the parameters of the LCA as a function of sample size (abscissa) and for different methods. More details can be found in the caption of Figure~\ref{mae}.}
     \label{rmseAll}
\end{figure}

\paragraph{Accuracy}
The recoveries of the original LCA parameters are displayed in Figures~\ref{N300}, \ref{N1000}, and \ref{N10000}. It can be concluded that for all sample sizes, recovery is acceptable, but it improves a lot for larger sample sizes.  In all cases, the recovery is dramatically better than that reported in \cite{miletic_parameter_2017}.
Figures~\ref{rmseAll} and~\ref{maeAll} shows RMSE and MAE, respectively, as a function of sample size for three methods (for all parameters). It can be seen that accuracy improves for all parameters for the single best nearest neighbor and for the bootstrap method, until some point, after which it stabilizes or deteriorates. However, for the SVM based estimation, there is still considerable improvement for higher sample sizes.

\begin{figure}[ht]
  \centering
    \includegraphics[width=0.5\textwidth]{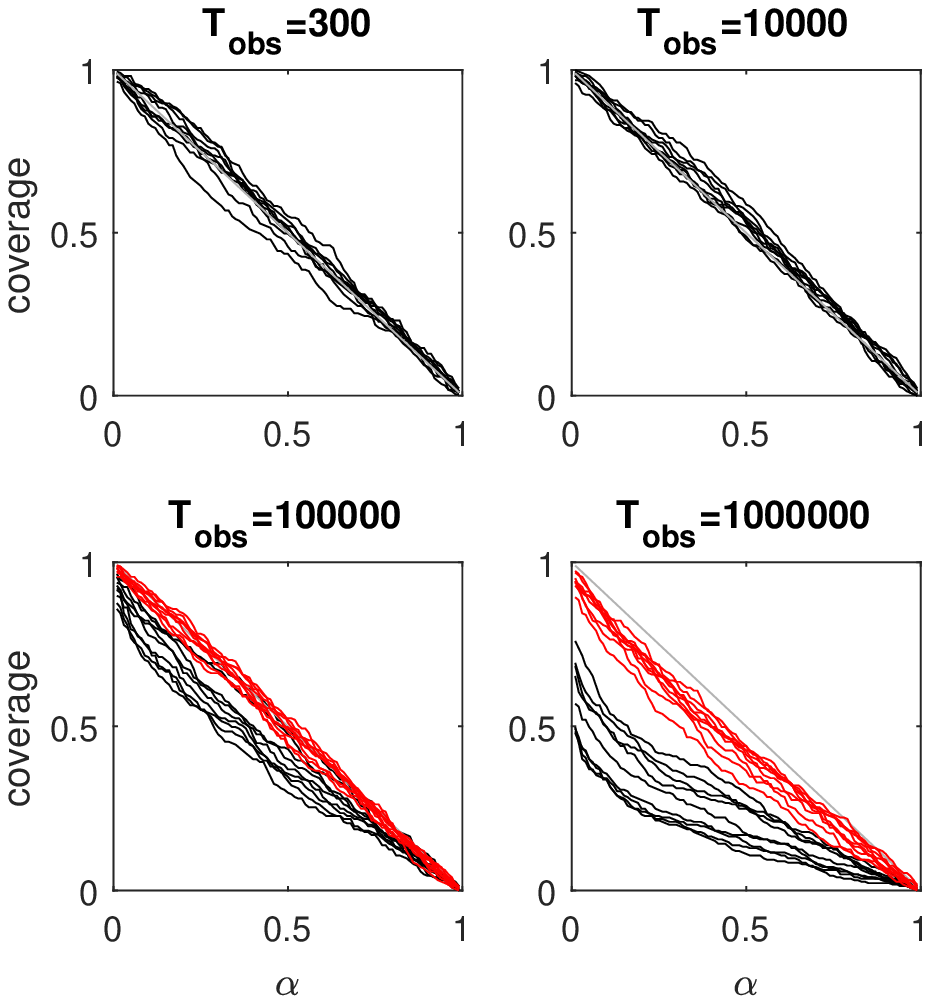}
     \caption{The coverage of LCA estimates for different number of observations $T_{obs}$. Each line represents one of the nine LCA parameters and plots the fraction of estimates between the $[\alpha,1-\alpha]$ quantiles of their bootstrapped confidence intervals. The closer the line to the second diagonal, the better the coverage. Black lines are the result of non-parametric bootstraps obtained through nearest neighbor estimates; red lines are the result of SVM enhanced estimates.} 
     \label{LCAcoverage}
\end{figure}

\paragraph{Coverage}
Figure~\ref{LCAcoverage} shows the coverages for different numbers of observations. Nearest neighbor bootstrap coverage seems to be adequate for sample sizes up to 10000; for higher sample sizes SVMs are needed to ensure good coverage.

% \begin{figure}[ht]
%   \centering
%     \includegraphics[width=0.5\textwidth]{LCAv7cB_v4_maxAcc99_N300_1sets_qsD_100nBs_scatter_all.eps}
%      \caption{Recovery for the original parameters of the LCA model with $T_{obs}=300$ observation per stimulus (true value on the abscissa and estimated value on the ordinate). The method used to produce these estimates is the single nearest neighbor: first, conditional on each grid point, estimating some subset of parameters in a preprocessing stage, and second, selecting the best fitting grid point based on Equation~\ref{eq:objfun_lca}. In total there are four stimuli. The four drift rates $v_1,\dots,v_4$ are taken together in the plot. The drift rates, $\gamma$ and $\kappa$ are displayed on a log-log scale for ease of interpretation (the drift rates are also presented on their original scale).}
%      \label{N300}
% \end{figure}

\section*{Acknowledgements}

The research leading to the results reported in this paper was sponsored in part by Belgian Federal Science Policy within the framework of the Interuniversity Attraction Poles program (IAP/P7/06), as well as by grant GOA/15/003 from the KU Leuven, and the grand for M.M and grant G.0806.13 from the Fund of Scientific Research Flanders.

 \section*{Author contributions statement}

 M.M. and S.V. conceived the method; M.M., S.V., and K.M. implemented the method. T.L. and F.T. studied the method theoretically. M.M., S.V., and F.T. wrote the manuscript.

\section*{Additional information}
The author(s) declare no competing interests.

\bibliographystyle{abbrvnat}
\bibliography{prepaid}

\end{document}